\documentclass[pdftex,twoside,11pt,letterpaper]{article}

\skip\footins=4ex						  
\usepackage{amsmath, amsfonts}
\usepackage[numbers,sort]{natbib}
\usepackage[pdftex]{graphicx}
\usepackage{caption}
\usepackage{subcaption}
\usepackage{wrapfig}
\usepackage{floatflt}
\usepackage{float}
\usepackage{calc}
\usepackage{color}
\restylefloat{figure}

\newcommand{\comment}[1]{}

\newcommand{\beginsupplement}{
        \setcounter{table}{0}
        \renewcommand{\thetable}{S\arabic{table}}
        \setcounter{figure}{0}
        \renewcommand{\thefigure}{S\arabic{figure}}
}

\begin{document}

\title{Network-based coverage of mutational profiles reveals cancer genes}
\author{Borislav H. Hristov and Mona Singh\footnote{Department of Computer Science and Lewis-Sigler Institute for Integrative Genomics, Princeton University}~\footnote{Email mona@cs.princeton.edu}}
\date{\vspace{-5ex}}

\maketitle
\thispagestyle{empty}

\begin{abstract}
A central goal in cancer genomics is to identify the somatic
alterations that underpin tumor initiation and progression. This task
is challenging as the mutational profiles of cancer genomes exhibit
vast heterogeneity, with many alterations observed within each
individual, few shared somatically mutated genes across individuals,
and important roles in cancer for both frequently and infrequently
mutated genes.  While commonly mutated cancer genes are readily
identifiable, those that are rarely mutated across samples are
difficult to distinguish from the large numbers of other infrequently
mutated genes.  Here, we introduce a method that considers
per-individual mutational profiles within the context of
protein-protein interaction networks in order to identify small
connected subnetworks of genes that, while not individually frequently
mutated, comprise pathways that are perturbed across (i.e., ``cover'')
a large fraction of the individuals.  We devise a simple yet intuitive
objective function that balances identifying a small subset of genes
with covering a large fraction of individuals.  We show how to solve
this problem optimally using integer linear programming and also give
a fast heuristic algorithm that works well in practice.  We perform a
large-scale evaluation of our resulting method, {\tt nCOP}, on 6,038
TCGA tumor samples across 24 different cancer types. We demonstrate
that our  approach {\tt nCOP} is more effective in identifying cancer genes than both
methods that do not utilize any network information as well as
state-of-the-art network-based methods that aggregate mutational
information across individuals.  Overall, our work demonstrates the
power of combining per-individual mutational information with
interaction networks in order to uncover genes functionally relevant
in cancers, and in particular those genes that are less frequently
mutated.\\

\noindent
{\bf Software download:} {\tt{http://www.cs.princeton.edu/$\sim$mona/software/ncop.html}}

\end{abstract}

\clearpage
\setcounter{page}{1}
\section*{Introduction}


Large-scale cancer genome sequencing consortia, such as The Cancer
Ge\-nome At\-las (TCGA)~\cite{tcga}, the International Cancer Ge\-nome Consortium
(ICGC)~\cite{ICGC} and other smaller, cancer-specific studies have sequenced the
protein-coding regions of thousands of tumor samples across tens of
different cancer types. Initial analyses of these data have
revealed that while there may be numerous somatic mutations in a
tumor that result in altered protein sequences, very few are likely to 
play a role in cancer development ~\cite{Bozic2010,Vogelstein2013,Garraway2013}.  Therefore, a major challenge in
cancer genomics is to develop methods that can 
distinguish the so-called ``driver'' mutations important for cancer
initiation and progression from numerous other ``passenger''
mutations.


Early statistical approaches have identified cancer-driving genes by
highlighting those genes that are mutated more frequently in a cohort
of patients than expected by chance according to some background
model~\cite{Youn2011,DeesZhKa12,lawrence2013mutational}. However, the genetic underpinnings of cancer are
highly heterogeneous: even when considering a single cancer
type, very few genes are found to be somatically mutated across large
numbers of individuals~\cite{hudson2010international}.  Further, genes
altered only in a few individuals may also be important for
tumorigenesis and cancer progression~\cite{Stratton2009}. Clearly, these rarely
mutated but cancer-relevant genes cannot be detected by purely
frequency-based approaches.

A promising alternative viewpoint is to consider somatic mutations in
the context of pathways instead of genes.  In particular, it has been
proposed that alterations within any of several genes comprising the
same pathway can have similar consequences with respect to cancer
development, and that this contributes to the mutational heterogeneity
evident across cancers.  Consistent with this, numerous analyses of TCGA
data have shown that certain known pathways are frequently altered
across tumor samples of a particular cancer via mutations in different
genes~\cite{TCGA-colorectal,mclendon2008comprehensive}. Early studies
have leveraged this observation by analyzing known pathways for
enrichment of somatic mutations~\cite{jones2008core,
  Netbox2010} and pinpointing those that are significantly
mutated across patients~\cite{pmid21498403,VaskeBeSa10}. The power of
these studies is somewhat limited, however, as our knowledge of
pathways is incomplete and new pathways cannot be identified by these
approaches.

\emph{De novo} discovery of cancer-relevant pathways using large-scale
protein interaction networks has thus been the focus of several newer
methods
(e.g.,~\cite{vandin2011algorithms,Netbox2010,memo2012,tiedie2013,cho2016muffinn}).
In particular, since protein-protein interaction networks have a
modular organization~\cite{HartwellHoLe99,SpirinMi03}, proteins taking
part in the same pathways and processes tend to be close to each other
in the network.  One prominent class of techniques leverages this modular structure by propagating mutational
information through protein interaction networks  and deriving pathways from the induced
subnetworks~\cite{vandin2011algorithms,Leiserson15,jia2014varwalker,babaei2013}. For
instance, Vandin et al.~\cite{vandin2011algorithms} diffuse a ``heat''
signal arising from the frequency with which proteins are somatically
mutated across a cohort of samples to uncover cancer-relevant modules
while Hofree et al.~\cite{hofree2013network} approach the problem from
a different angle, using biological network information to stratify
cancer subtypes. A recent pan-cancer network
analysis~\cite{Leiserson15} affirms the power of diffusing 
mutational data across protein interaction networks, especially for
uncovering rarely mutated cancer genes. However, such diffusion
approaches can be highly influenced by frequently mutated
genes~\cite{Leiserson15}, and further, these methods do not consider whether
most patients have mutations in any of the identified pathways.





Here we present a novel network-based approach to tackle cancer
mutational heterogeneity by utilizing per-individual mutational
profiles.  Our method is based on the expectation that if a pathway
is relevant for cancer, then (1) many individuals will have a somatic
mutation within one of the genes comprising the pathway and (2) the
genes comprising the pathway will interact with each other and
together form a small connected subcomponent within the larger
network.  Therefore, given a biological network as well as patient sample data
consisting of somatic point mutations, the goal of our
approach is to find a set of candidate genes that both
``cover'' the most patients (i.e., individuals have mutations in one
or more of these genes) and are connected in the network (i.e., these genes are likely
to participate in the same cellular pathway or process).  In contrast
to network diffusion approaches, our framework focuses on
per-individual mutational profiles and as a result, the ``influence''
of frequently mutated genes is not spread through the network. We note
that network-based coverage approaches have been previously introduced
to uncover pathways that are
dysregulated~\cite{ulitsky2010degas,ChowdhuryKo10,KimWuPr11} or
mutated~\cite{dand2013biogranat,kim2015memcover} across cohort of
samples. However, either patients were required to be covered by these
approaches~\cite{ulitsky2010degas,ChowdhuryKo10,KimWuPr11,kim2015memcover},
in some cases multiple times (which is especially relevant for
dysregulated genes, since there many of them), or these approaches
were designed for data sets with significantly fewer
mutations~\cite{dand2013biogranat}; both cases lead to very different
optimizations and algorithms that are not effective for the task at
hand.


We devise a simple yet intuitive objective function that
balances identifying a small subset of genes with covering a large fraction of
individuals.  Our objective has just a single parameter
that is automatically set using a series of cross-validation tests,
eliminating the need of many previous approaches to manually select
values for various thresholds and parameters.  We develop an integer
linear programming formulation to solve this problem and also give a 
fast heuristic algorithm.
We apply our method---\textbf{n}etwork-based \textbf{co}verage of
\textbf{p}atients ({\tt nCOP})---to 24 cancer types from TCGA
and uncover both well-known cancer driver genes as well as new
potential cancer-related genes. We compare {\tt nCOP} to previous
methods that do not use network information, including a
state-of-the-art frequency-based method~\cite{lawrence2013mutational} and a 
``set cover'' version of our approach that attempts to find a set of genes that covers cancer
samples without considering network connectivity, and demonstrate 
{\tt nCOP}'s superior power in detecting known cancer genes and in zooming in on
rarely mutated ones. Finally, we compare {\tt nCOP} to a recent
network-based method that aggregates mutational
information~\cite{cho2016muffinn} and show that our per-patient
approach readily outperforms it.

\vspace{-.1in}
\section*{Methods}

\vspace{-.1in}

{\bf Overview.}  In this section, we give an overview of our
methodology (see also Supplementary Figure~1); each part is described in more
detail in the subsequent sections.

The biological network is modeled as an undirected graph $G$ where each
vertex represents a gene, and there is an edge between two vertices if
an interaction has been found between the corresponding proteins.  We
annotate each node in the network with the IDs of the individuals
having one or more mutations in the corresponding gene. Our goal is to
find a relatively small connected component $G'$ such that most
patients have mutations in one of the genes within it. A small subgraph is more likely to consist of
functionally related genes and is less likely to be the result of
overfitting to the set of individuals whose diseases we are
analyzing. However, we would also like our model to have the
greatest possible explanatory power---that is, to account for, or
cover, as many patients as possible by including genes that are
mutated within their cancers.  We formulate our problem to balance
these two competing objectives with a parameter $\alpha$ that controls
the trade-off between keeping the subgraph small and covering more
patients.  

For a fixed value of $\alpha$, we have developed two approaches to
solve the underlying optimization problem. One is based on linear
programming and the other is a fast greedy heuristic (see below).  We
use the greedy heuristic in the context of a carefully designed cross-validation
procedure to automatically select a value for $\alpha$ that results in good coverage of patients but avoids
overfitting to them.   Once $\alpha$ is
selected, this value is used within our objective function and we 
next analyze the  entire patient cohort. In particular,
multiple independent trials using $\alpha$ are run on randomly chosen subsets of the
patient data, as we have found that introducing a little bit of
randomness helps increase performance as compared to a single run on the
full data set. Each trial outputs a subgraph, and our final aggregated
output is an ordered list of candidate genes ranked by how frequently
each has been selected over the trials.

\smallskip
\noindent
{\bf General formulation.}  Each vertex $v_j$ is associated with a set
$C_j$ containing the IDs of the individuals who have somatic mutations
in the corresponding gene. A patient with ID $i$ is covered if $i \in
\bigcup\limits_{v_j \in G'} C_j$, and uncovered otherwise. We
formulate our problem as that of finding a connected subgraph $G'$ of
$G$ so as to minimize $$\alpha X + (1-\alpha) Size(G'),$$ where $X$ is
the fraction of patients that do not have an alteration in a gene
included in $G'$ (i.e., they are uncovered), $Size(G')$ is the size of
the subgraph, and $0 \le \alpha \le 1$ is a fixed parameter
controlling the trade-off between keeping the subgraph $G'$ small and
covering more patients.  We note that our problem is similar, though
not identical, to the Minimum Connected Set Cover Problem
\cite{shuai2006connected}, a NP-hard problem.

A simple and natural measure for the size of a subnetwork is its
number of nodes (i.e., $Size(G')=|G'|$). However, longer genes may
tend to acquire more mutations simply by chance. We correct for that
by associating with each node $v_j$  a weight $w_j$ that is equal to the
ratio of the length of the gene to the total number of mutations it
has. The size of the subcomponent is then defined as
$Size(G')=\sum\limits_{v_j\in G'}w_j$. This way, genes having longer
length will be weighted more, correcting for a possible bias towards
selecting longer genes.  We note that since our objective function
balances the fraction of uncovered patients with the size of the
graph, we would like the  size of the graph to be between 0 and
1; thus, we normalize each node weight by dividing by the unnormalized size of what
we call a fully covering subgraph $G^f$---a connected subgraph of $G$
that covers all patients. (In practice, we compute $G^f$ using the
greedy heuristic described below, with $\alpha=1$).

\smallskip
\noindent
{\bf Integer linear programming formulation.}  
The problem of finding a minimum connected subgraph that covers as
many patients as possible can be solved using constraint optimization. 
Let $n$ be the number of patients in our sample.  For each patient
$i$, we define a binary variable $p_i$ that is set to~1 if
patient $i$ is covered by the chosen subgraph $G'$, and 0 otherwise.
For each vertex (or gene) $v_j$, we define a binary variable $x_j$
that is set to~1 if the vertex is included in the chosen subgraph
$G'$, and 0 otherwise.  It is straightforward to set up constraints to
ensure that a patient is considered uncovered if none of its mutated
genes are part of $G'$, and covered if at least one of its mutated
genes is selected as part of $G'$ (see Equations (1) and (2) below).

The challenging part of the ILP is setting up constraints to ensure
that the chosen nodes form a connected subgraph $G'$.  For this task,
we employ a flow of commodity technique~\cite{even1975network}, which
we now briefly describe. We inject $|G'|$ units of flow into $G'$
(i.e., we inject $\sum x_i$ units of ``flow'' into a vertex that is
included in the chosen subnetwork). Flow can move from one vertex to
any of its neighbors in the network, and each vertex removes exactly
one unit of flow as the flow passes through it.  All flow must be
removed from the subnetwork, and we set the constraints so that this
is possible only if the subnetwork $G'$ is connected. For the source
of the flow we use an artificial external node $v_{extr}$.  The main
issue is that we do not know which node $v_{extr}$ should be connected
to, as we do not know the nodes of $G'$ in advance. To resolve this,
we decide that $v_{extr}$ connects to the node that covers the largest
number of patients $v_{max}$; this is equivalent to determining in
advance that $v_{max}\in G'$, though as an alternate approach we could
also decide to choose this node probabilistically and run the ILP
several times.  Finally, to handle the flow constraints, for each edge
$(i, j) \in E$, we introduce integer variables $y_{i,j}$ and $y_{j,i}$
to represent the amount of flow from node $i$ to node $j$ and from
node $j$ to node $i$, respectively. The full integer linear program
is:

\newpage
\noindent\emph{minimize }$\alpha(n-\sum\limits_i p_i)/n + (1-\alpha) \sum\limits_j x_jw_j$\\
\noindent\emph{subject to}\\
\begin{align}
p_i & \ge   x_j &  \forall i, j \text{~s\@. t.~}  i\in C_j \\  %
p_i & \le   \sum\limits_{j: i \in C_j} x_j & \text{for each patient $i$} \\
\sum\limits_{i: (i, j) \in E}  y_{i,j} &= x_j + \sum\limits_{i: (i, j) \in E} y_{j,i} &  \text{for each vertex $v_j$}\\
\sum\limits_{j: (i,j) \in E} y_{i,j} &\le |V|x_i  & \text{for each vertex $v_i$}\\
\sum_i x_i &= y_{extr,max}\\
p_i, x_i, y_{i,j} & \in  \{0, 1\} &  \text{for all such variables}
\end{align}
Equation (1) ensures that a patient is considered
  covered if one of his or her or somatically mutated genes is
  included in $G'$. Equation (2) ensures that a patient is not
  considered covered if none of his or her somatically mutated genes
  is chosen to be part of the subgraph.  Equations (3), (4) and (5)
  enforce the connectivity requirement. Equation~(3) requires that the
  flow going out of each vertex in the chosen subnetwork is~1 less than 
  the flow coming in.  Equation~(4) requires that if a vertex
  is not part of the chosen subgraph, the flow going through it is 0.
  Equation~(5) sets the amount of flow injected into the subgraph to
  be equal to the number of chosen nodes.


\smallskip
\begin{sloppypar}
\noindent
{\bf Greedy heuristic.}  Solving the ILP yields an exact solution but
is computationally difficult. Thus, we have also developed an
efficient greedy heuristic.  Our heuristic procedure initializes $G'$
by randomly choosing the first gene from among the five most
mutated genes, with probability proportional to the number of patients it is found mutated
in.  It then expands the subgraph $G'$ iteratively as follows.  At
each iteration, all vertices that are at most distance~2 from a vertex
in $G'$ are examined and the one that improves the objective function
the most is chosen; any ties are broken uniformly at random.  If this
vertex is not directly adjacent to the nodes in the subnetwork, the
intermediary node is also added.  The heuristic terminates when no
improvement to the objective is possible.  We repeat this heuristic
multiple times, as it is probabilistic.
\end{sloppypar}

In practice, the greedy heuristic finds a solution that is on average
$\sim$90\% of the best value for the objective function as determined
by the ILP formulation using {\tt CPLEX}~\cite{CPLEX}. For example, on the
glioblastoma dataset of 277 individuals, the ILP
finds 61 genes covering 90\% of the patients when using
 $\alpha$=0.5.  In comparison, for this value of $\alpha$, the greedy
heuristic finds on average 66 genes covering 88\% of the patients. In the
rest of the paper, we use the greedy optimization as it has
comparable performance to the ILP, while being much faster.

\smallskip
\noindent
{\bf Parameter selection and solution aggregation.} Our approach to
uncover a subnetwork of mutated genes that covers many patients has a
single parameter, $\alpha$.  Large values of $\alpha$ result in a
larger number of selected genes that cover more patients, yet may contain more
irrelevant genes; this may especially be a factor in the current analysis if there are many
samples where missense mutations are not the driving event.  We devised a
simple but effective data-driven cross-validation technique to choose
an appropriate $\alpha$ for a set of cancer samples.  In particular,
we split our samples into training, validation and test
sets~\cite{hastie2009elements}. A test set of (10\%) of the patients
is completely withheld. While varying $\alpha$ in small increments in
the interval $(0;1)$, the remaining data is repeatedly split (100
times for each value of $\alpha$) into training (80\%) and validation
(20\%) sets. For each split, the greedy heuristic algorithm is run on
the training set to find $G'$. The fractions of patients covered (by
the selected $G'$) in the training and validation sets are
compared. The parameter $\alpha$ is selected where performance on the
validation sets deviates as compared to the training sets. While this
can be done visually, for all results reported here we do this
automatically using a simple two-rule procedure that selects the
smallest $\alpha$ for which the difference in average coverage between
the training and validation set exceeds $5\%$ and for which average
performance on the validation set is within $10\%$ from the maximum
observed one for any $\alpha$. Finally, the coverage of patients on
the (completely withheld) test set is computed to ensure it is similar
to the one on the validation set.

Once $\alpha$ is chosen for a set of cancer samples, we repeatedly
(1000 times) run the algorithm on this set, each time withholding a
fraction ($15\%$) of the patients in order to introduce some
randomness in the process.  Genes are then ranked by the number of
times they appear in $G'$.  In practice, we have found that this
improves performance as compared to running the algorithm once on the
full data set.

\smallskip
\begin{sloppypar}
\noindent
{\bf Data sources and pre-processing.}  We downloaded all level 3
cancer somatic mutation data from The Cancer Genome Atlas
(TCGA)~\cite{tcga} that was available as of October 1, 2014. This data
consists of a total of 19,460 genes with somatic point mutations
across 24 cancer types.  For each cancer, samples that are obvious
outliers with respect to their total number of mutated genes are
excluded. See Supplementary Table 1 for a list of the cancer types, the cancer-specific thresholds
to determine outlier samples, the number of patient samples considered
for each cancer type, and other statistics about the TCGA somatic
mutation dataset.

We use two different biological networks in our analysis:
\emph{HPRD} \cite{prasad2009human} (Release 9\_041310) and \emph{BioGrid}
(Release 3.2.99, physical interactions only)~\cite{stark2006biogrid}. 
Biological networks can exhibit several nodes with very high connectivity,
often due to study bias. As such high connectivity destroys the
usefulness of the network information, we remove all nodes whose
degrees are clear outliers. For \emph{BioGrid}, this removes
\emph{UBC, APP, ELAVL1, SUMO2, CUL3}. For \emph{HPRD}, this removes no nodes. For both
networks, we exclude the nine longest genes (\emph{TTN, MUC16, SYNE1,
  NEB, MUC19, CCDC168, FSIP2, OBSCN, GPR98}) as they tend to acquire
  numerous mutations by chance while covering many patients.

To further handle the connectivity arising within the
networks due to high-degree nodes, we filter edges using the diffusion
state distance (DSD) metric introduced in~\cite{cao2013going}; the DSD
metric captures the intuition that edges between nodes that also share
interactions with low degree nodes are more likely to be functionally
meaningful than edges that do not (and thus are assigned closer
distances).    For each edge, the DSD scores (as computed by the software of~\cite{cao2013going})
between the corresponding nodes are Z-score normalized, and edges with Z-scores
$>0.3$ are removed.  We note that the overall performance of our
approach improves when performing this filtering (data not shown),
supporting the claim of \cite{cao2013going} that preprocessing a
biological network in this manner is an important step.  The final number of nodes
and edges, respectively, for the filtered networks are 9,379 and
36,638 for \emph{HPRD}; and 14,326 and 102,552 for \emph{BioGrid}.
\end{sloppypar}

\smallskip
\noindent
{\bf Other approaches.} To ascertain the contribution of network
information, we compare {\tt nCOP} to two approaches that do not use
network information: (1) {\tt MutSigCV
2.0}~\cite{lawrence2013mutational}, a state-of-the-art method that
identifies genes that are mutated more frequently than expected
according to a background model, and (2) a set cover approach that
tries to find mutated genes that simply cover as many patients as
possible.  We formulate the set cover approach as an ILP that tries to
find a good cover consisting of $k$ vertices.  Using the same notation as for {\tt
nCOP}, the set cover objective is to \emph{maximize }$\sum\limits_i
p_i$, subject to Equations (1) and (2) of {\tt nCOP}, and with the
additional constraints that $\sum\limits_j x_j \le k$ and
$\sum\limits_j x_j \ge k$.  We also compare {\tt nCOP} to
HOTNET2~\cite{Leiserson15} and {\tt Muffinn}~\cite{cho2016muffinn},
two recent network-based approaches that aggregate mutational information.  All comparisons are made when
running on exactly the same cancer and (if used) network data.  {\tt
MutSigCV 2.0}, {\tt Hotnet2} and {\tt Muffinn} are run with default
parameters.



\smallskip
\noindent
{\bf Performance evaluation.} To evaluate the gene rankings of
all the tested methods, we use the curated list of 517 cancer census genes
(CCGs) available from COSMIC~\cite{Futreal2004}. All genes in this list
are considered as positives, and all other genes are considered as
negatives.  Though we expect that there are genes other than those already on
the CCG list that play a role in cancer, this is a standard approach to
judge performance (e.g.,see~\cite{jia2014varwalker}) and gives us an idea
of how methods are performing as cancer genes should be highly ranked
by methods that perform well.  Since only the top predictions by any
method are relevant for cancer gene discovery, we judge performance by
computing the area under the precision-recall curve (AUPRC) using the
top 100 genes predicted by each method. If a method returns less than
100 genes, we extend the precision-recall curve to 100 genes assuming that
it performs as a random classifier. We note that reasonable changes to
the number of predictions considered does not change our overall conclusions (data not shown).

\vspace{-.1in}
\section*{Results}
\vspace{-.1in}
We run nCOP, using the greedy heuristic algorithm, on somatic point mutation
data from 24 different TCGA cancer types.
Results in the main paper use the \emph{HPRD}
network~\cite{prasad2009human} for all analysis and show some results using kidney renal clear cell
carcinoma (KIRC) with 416 samples as an examplar.


\smallskip
\noindent
{\bf Increasing patient coverage while preventing overfitting.} 
We first demonstrate that, across the 24 cancer types,  our training-validation-test set framework
is a highly effective approach for choosing an $\alpha$ that balances
patient coverage with subnetwork size.  
For all cancers, when using somatic missense mutation data, as $\alpha$ increases, the total number
of genes in the chosen subnetwork $G'$ increases (as expected), as does the fraction
of patients in the training set that are covered by these genes
(Figure 1a and Supplementary Figure S2). For smaller values of
$\alpha$, coverage on the validation sets closely matches that obtained on the
training sets; that is, the sets of genes chosen using patients in the
training sets are also effective in covering patients in the corresponding validation
sets.  For KIRC, when $\alpha = 0.5$, genes chosen using the training sets
cover on average nearly 70\% of patients in the corresponding validation sets. The fact that a
small subnetwork can be found that covers a large fraction of previously unseen patients
is consistent with the hypothesis that a shared pathway or process
plays a role in most (but not all) of these patients' cancers.

\begin{figure}[t!]
	\begin{subfigure}{.49\textwidth}
	  \includegraphics[width=.95\linewidth]{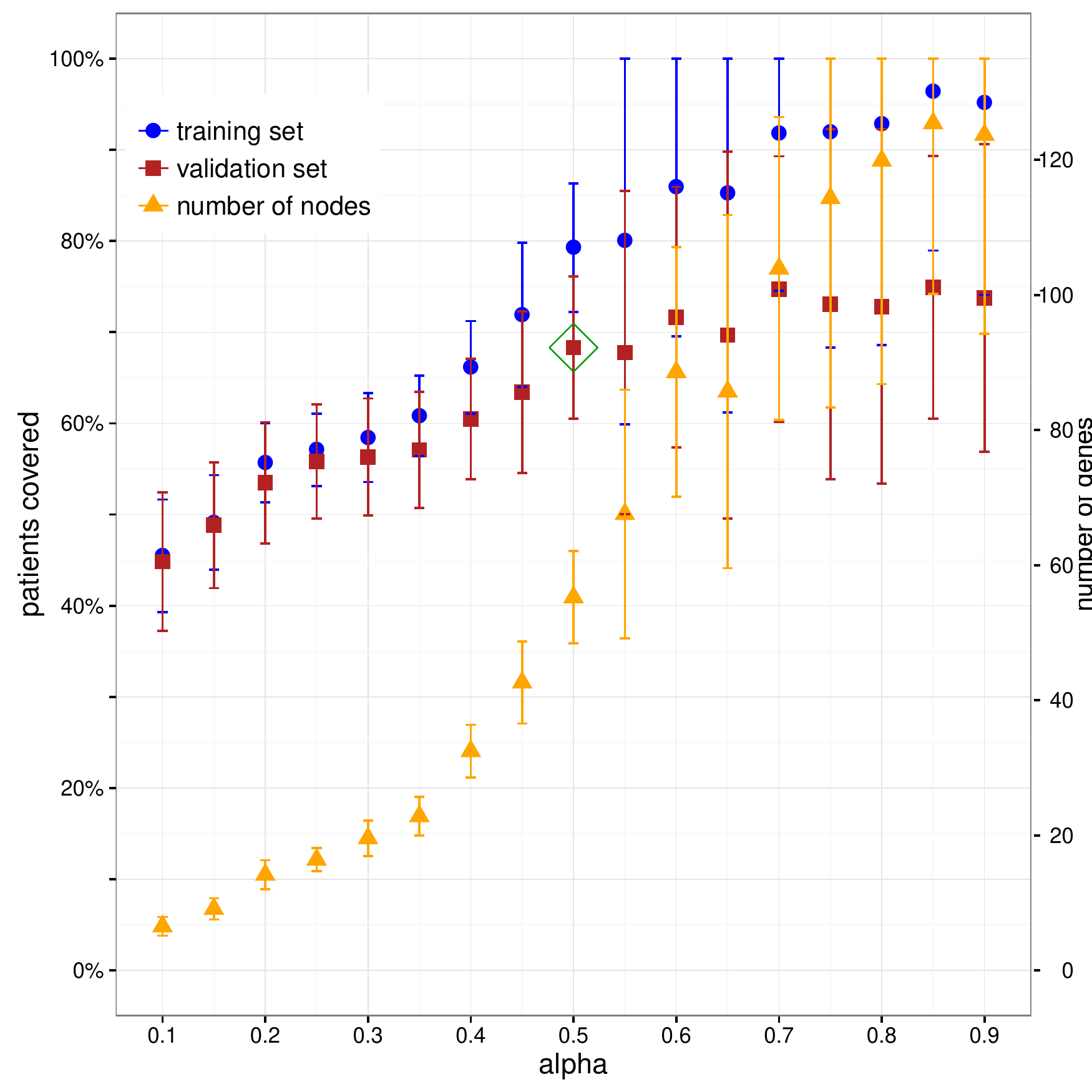}
	  \caption{missense mutations}
	\end{subfigure}
	\begin{subfigure}{.49\textwidth}
	  \includegraphics[width=.95\linewidth]{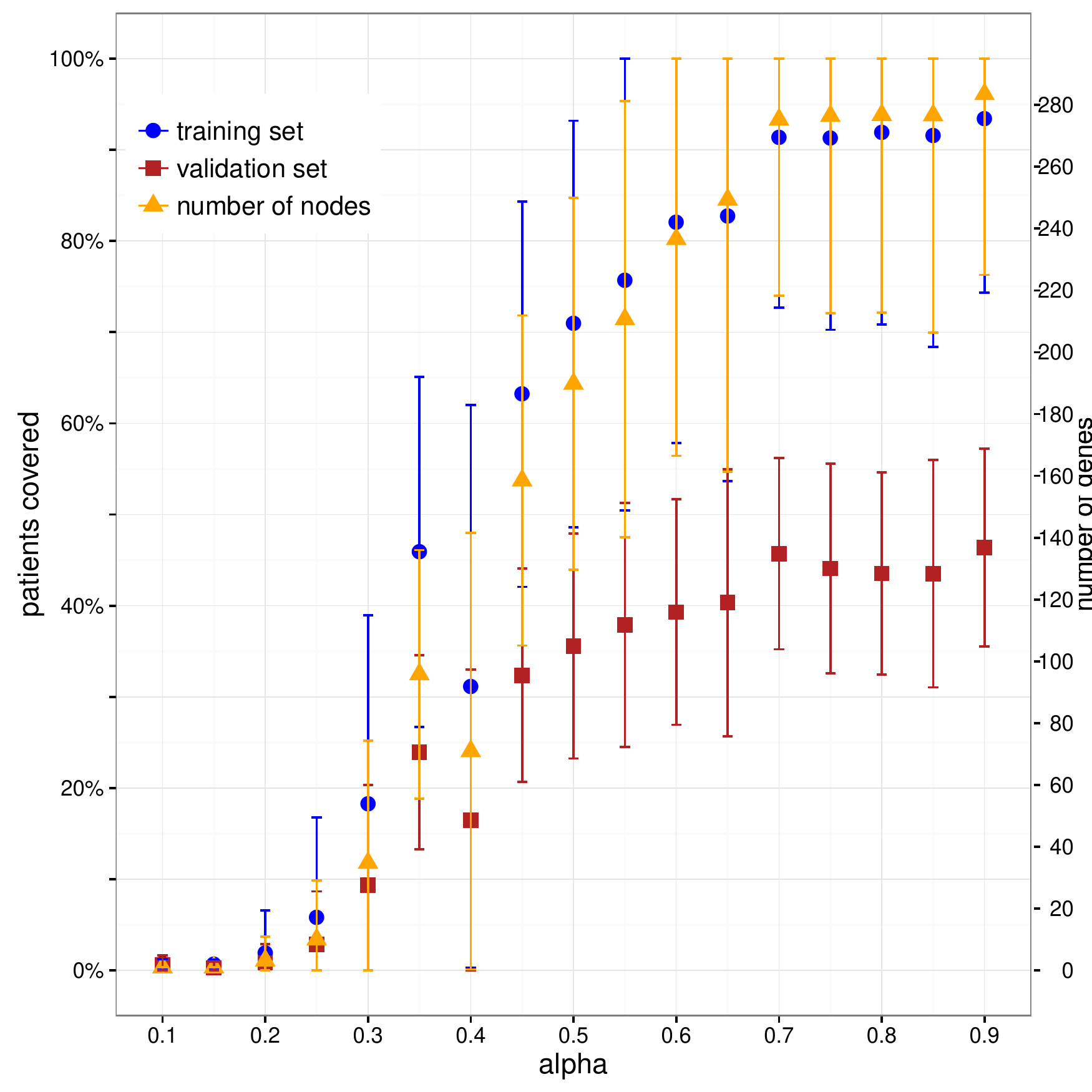}
	  \caption{synonymous mutation}
	\end{subfigure}
	
	\caption{\small {\bf Fraction of individuals covered as $\alpha$
            varies.}  We illustrate our cross-validation procedure for
          parameter selection using the 
          KIRC data set and the HPRD protein-protein
          interaction network.  For each random split of the
          individuals, we run our algorithm on the training sets for
          different values of $\alpha$, and plot the fraction of
          covered individuals in the training (blue) and validation
          (red) sets.  We also give the number of proteins in the
          uncovered subgraphs $G'$ (orange). For each plotted value, the mean
          and standard deviation over 100 random splits are
          shown. {\bf (a)} When using somatic missense mutations, at
          higher values of $\alpha$, overfitting occurs as the
          coverage on the validation set levels while coverage on the
          training set continues to increase. The parameter $\alpha$
          is selected using an automated heuristic procedure (green
          rhombus) so that coverage on the validation set is good
          while overfitting on the training set is not extreme.  {\bf
            (b)} When using somatic synonymous mutations, there is
          poor coverage on the validation set regardless of coverage
          on the training set. Further, as compared to using missense
          mutation data, significantly more genes are required to
          cover the same fraction of individuals.}
\end{figure}

For larger values of $\alpha$ ($> 0.6$ for KIRC), however, coverage on
the validation sets lags behind that observed on the training
sets. For even larger values of $\alpha$ ($> 0.85$ for KIRC), the
algorithm selects many genes, and eventually increases the coverage
for most cancers on the training sets to nearly 100\%. However, larger
values of $\alpha$ do not substantially increase coverage of the
withheld patients. This difference between the training and validation
curves captures the overfitting of the model and also illustrates the
trade-off between covering more patients and keeping the solution
parsimonious. We note that the eventual plateau of the validation
curve is consistent across cancer types (Supplementary Figure S2) and
on different networks (data not shown).  For each cancer type, values
of $\alpha$ are selected by our automated procedure (see Methods);
this value is $\alpha = 0.5$ for the KIRC dataset shown in Figure 2a.

As a control, we repeat the same procedure using only synonymous
mutations (Figure 1b).  We observe that the coverage on the validation
sets is much poorer. Though coverage of course increases as more nodes
are added, it never exceeds $50\%$ even when $\alpha$ is increased to
$1$ or when we have nearly perfect coverage on the training set,
despite adding many more nodes. This poor performance is consistent
with the expectation that synonymous mutations do not result in
altered protein sequences and do not disturb cellular pathways.
Hence, given the differences observed between using missense versus
silent mutation data when varying settings for $\alpha$ and comparing
training and validation sets, our formulation appears to be
well-suited for extracting cancer-relevant information from mutational
profiles and interaction networks.

\smallskip
\noindent
{\bf nCOP effectively uses network information to uncover known cancer genes.} 
Having shown in the previous section how to select a value for the
only parameter in the model and having demonstrated that our formulation
is effective in choosing genes that are mutated in previously
unobserved patient samples, we next evaluate {\tt nCOP}'s performance
in predicting cancer genes.

We first consider how well  {\tt nCOP} recapitulates
known cancer genes (CCGs)~\cite{forbes2010cosmic}. We find that on KIRC, our top
predictions include a high fraction of CCG genes (Figure~2).  For example, genes that
are always output are \emph{VHL}, \emph{PTEN}, and \emph{BAP1}, three well-known cancer
genes. Other known cancer genes such as \emph{JUN}, \emph{BLM} and
\emph{ARID1A}, are also highly ranked.


We next illustrate the power of our network-based method by comparing
its performance on the KIRC data set  to approaches that do not consider network information
(Figure~2). First, we consider a set
cover version of our approach that does not use network information
at all. We find that for
the same number of predicted genes, our approach consistently has a
larger fraction of CCGs, demonstrating the advantage of using network
information. We also considerably outperform a state-of-the-art
frequency-based approach, {\tt MutSigCV~2.0}~\cite{lawrence2013mutational}.

\begin{floatingfigure}[ht]{0.5\linewidth}
\includegraphics[width=0.45\textwidth]{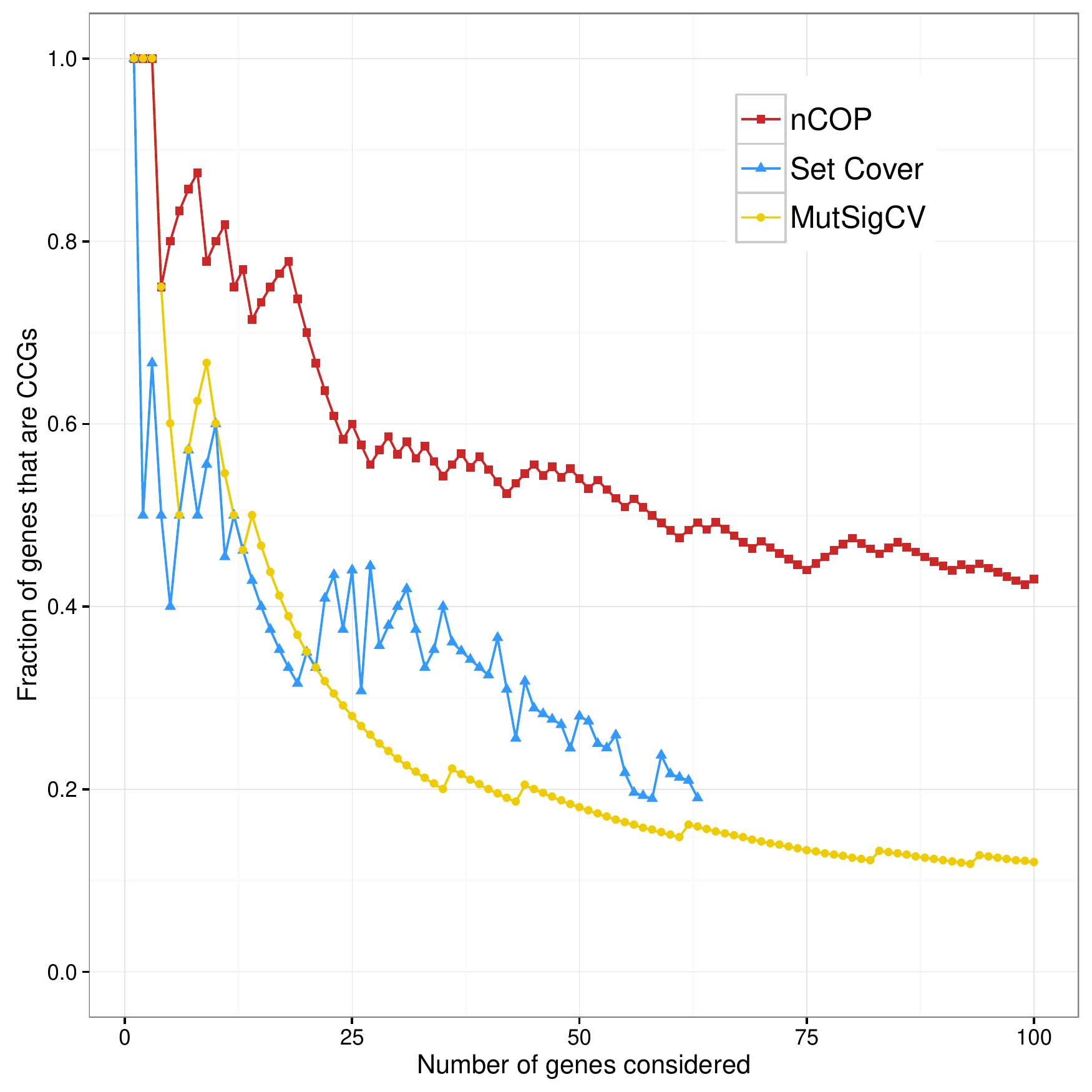}
\caption{\small {\bf {\tt nCOP} is more successful than network-agnostic methods in identifying
    known cancer genes on the KIRC dataset.}  Our
  network-based algorithm {\tt nCOP}, a set cover version of our
  algorithm that ignores network information, and {\tt MutSigCV~2.0}, a
  frequency-based approach, are compared on the KIRC dataset.
  {\tt nCOP} ranks genes based on how frequently they are output, and {\tt MutSigCV~2.0} ranks genes by $q$-values.  The set
  cover approach is run for increasing values of $k$ until all
  patients are covered.  For each method, as an increasing number of
  genes are considered, we compute the fraction that are CCGs. Over a
  range of thresholds, our algorithm {\tt nCOP} outputs a larger
  fraction of CCG genes than the other two approaches.  }
\end{floatingfigure}

\begin{sloppypar}
We next compare {\tt nCOP} to these two non-network approaches across
all 24 cancer types.  In particular, we compute the $\log$ ratio of
the area under the precision recall curve (AUPRC) of our approach
versus each of the other approaches on each cancer type (Figure~3). We
outperform {\tt MutSigCV~2.0} in 21 of the 24 cancers and the set cover approach
in all cancers, demonstrating the clear advantage of using network
information; the performance improvement of {\tt nCOP} over the set
cover approach is particularly notable as the main difference of these
approaches is the additional use of network information by
{\tt nCOP}.  In several cancers, the performance improvements of
{\tt nCOP} are substantial. For example, {\tt nCOP} shows a
four-fold improvement over {\tt MutSigCV~2.0} in predicting cancer genes for
liver hepatocellular carcinoma (LIHC) and a nearly eight-fold improvement over {\tt MutSigCV~2.0} on pheochromocytoma and
paraganglioma (PCPG).  Importantly, {\tt nCOP} uncovers rarely
mutated CCGs genes that network-agnostic methods fail to. For
instance, \emph{KLF6} and \emph{TCF7L2}, mutated respectively in only one and two
individuals in {LIHC}, and \emph{WT1}, mutated in two individuals in PCPG, are
all uncovered by {\tt nCOP} because of their proximity to other
mutated genes in the network. In contrast, frequency-driven {\tt MutSigCV~2.0}
and network-agnostic set cover both fail to uncover these CCG genes.  We
note that the suboptimal performance of {\tt nCOP} in the case of
uterine carcinosarcoma (UCS) is due to the fact that for this
particular data set, two genes \emph{TP53} and \emph{FBXW7} cover $85\%$ of the
patients, with only four genes being sufficient to cover $95\%$ of the
patients. This renders {\tt nCOP} incapable to effectively leverage
network information as it returns only four genes. Nevertheless,
the overall superior performance of {\tt nCOP} as compared to these
two non-network based approaches on the vast majority of cancers demonstrates its considerable power.
\end{sloppypar}


Having shown that {\tt nCOP} better identifies cancer-relevant genes
than two approaches that do not use network information,  we next
consider whether the specific way in which {\tt nCOP} uses network
information is beneficial. Towards this end, we compare the
effectiveness of {\tt nCOP} in uncovering cancer genes to {\tt
Muffinn}~\cite{cho2016muffinn}, a method published within the last
year that considers mutations found in interacting genes.   We find that in 20 out of
the 24 cancer types, {\tt nCOP} outperforms {\tt Muffinn} (Figure~4).
We also compare {\tt nCOP} to {\tt Hotnet2} ~\cite{Leiserson15}, a
cutting-edge network diffusion method. As {\tt Hotnet2} does not
output a ranked list of genes, we could not compute the area under the
precision recall curve. Instead, examining the complete list of genes
highlighted by both methods, we observe that {\tt nCOP} exhibits
significantly better precision while trailing slightly in recall
(Supplementary Figure S3).

\smallskip
\begin{sloppypar}
\noindent
{\bf nCOP reveals novel genes, including those that are rarely mutated.} 
In addition to ranking known cancer genes highly, {\tt nCOP} also
gives high ranks to several non-CCG genes that may or may not be
implicated in cancer, as our knowledge of cancer-related genes is
incomplete. We observe that non-CCG that are highly ranked by
{\tt nCOP} tend to be less frequently mutated. For example, among these novel 
predictions for KIRC are \emph{SALL1}, \emph{NR5A2}, and \emph{UBE2A} which have all recently been
implicated to play a role in cancers \cite{wolf2014vivo,lin2014lrh1,yu2016brd7}. 
\end{sloppypar}

\begin{floatingfigure}[t]{0.55\linewidth}
\includegraphics[width=0.5\textwidth]{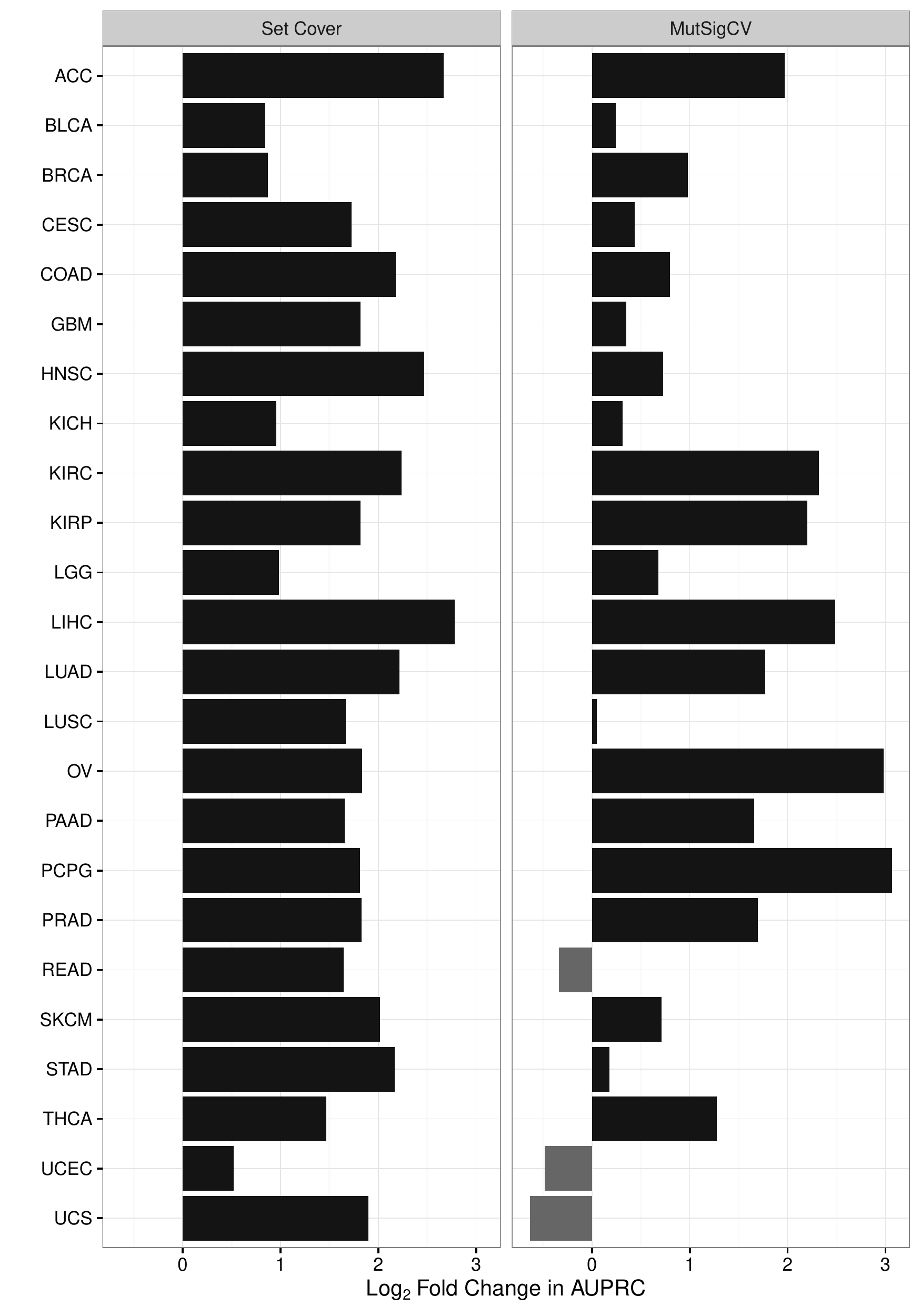}
\caption{\small {\bf Comparison of {\tt nCOP} to two network-agnostic methods across 24 cancer types.}  
                    For each of the 24 cancer types, we compute AUPRCs
			 for {\tt nCOP}, the set cover approach, and
			 {\tt MutSigCV 2.0} using their top 100
			 predictions. We give the $log_2$ ratios of
			 {\tt nCOP}'s to set cover's AUPRCs (left
			 panel) and of {\tt nCOP}'s to {\tt
			 MutSigCV~2.0}'s AUPRCs (right panel). Our
			 approach {\tt nCOP} outperforms the set cover
			 approach on all 24 cancers, and outperfoms
			 both network-agnostic methods on 21 out of 24
			 cancer types.}
\end{floatingfigure}

\begin{sloppypar}
When we consider the proteins output by our procedure in at least half
the runs for KIRC, we find that they are enriched in many KEGG and GO pathways
relevant for cancer, including \emph{MicroRNAs in cancer},
\emph{Choline metabolism in cancer}, \emph{Central carbon metabolism
  in cancer} and \emph{PI3K-Akt signaling pathway} (Bonferroni-corrected $p < 0.001$,
hypergeometric test).  When we remove all known
CCG genes and consider only new predictions, thus performing a much
harder test, only one pathway, \emph{Thyroid hormone signaling
  pathway}, is enriched.  Interestingly, it has been shown that thyroid
hormones play a role in kidney growth and
development~\cite{katz1975thyroid} and four of our non-CCG predictions
are part of that pathway, together with four known cancer
genes. Interestingly, two of the non-CCG genes we find in this pathway are
rarely mutated, and are not highlighted by {\tt MutSigCV~2.0} or set
cover. This illustrates the power of {\tt nCOP} to zoom in on
potentially cancer-relevant modules consisting of rarely mutated
genes.
\end{sloppypar}

\smallskip
\begin{sloppypar}
\noindent
{\bf Potential biases, performance on randomized data, and
  robustness.}  Due to space and time constraints, we briefly describe
  some additional tests we performed to show that {\tt nCOP} is robust
  and well-behaved. First, we have tested {\tt nCOP} on another
  network and have shown that it remains effective in identifying CCGs
  (Supplementary Figures S4).  Second, to confirm the importance of
  network structure to {\tt nCOP}, we have run {\tt nCOP} on two types
  of randomized networks, degree-preserving and label shuffling, and
  have shown that (as expected) overall performance deteriorates
  across the cancer types (Supplementary Figure S5); we note that
  though network structure is destroyed, per-gene mutational
  information is perserved, and so highly mutated genes are still
  output.  Third, to make sure that the novel genes we uncover are not
  driven by patients with large numbers of passenger mutations (i.e.,
  that the novel genes are not likely to be passenger genes), we
  have compared the overall number of mutations for patients having
  missense mutations only in CCG genes but not in any predicted non-CCG (or
  novel) genes to the total number of such mutations for patients having
  missense mutations only in novel genes but not in any CCG genes
  (Supplementary Figure S6), and have found that patients with only
  mutations in novel genes do not harbor more mutations.
   Finally, to make sure that genes are not more likely to be
  picked because they have higher degree, we have confirmed that newly
  predicted do not tend to exhibit higher degree than known cancer
  genes; indeed, among all novel genes found across all cancer types,
  most have degree between $<15$, and there are only a couple with
  high degree ($\ge 50$).
\end{sloppypar}

\vspace{-.15in}
\section*{Discussion}
\vspace{-.1in}
In this paper, we have shown that {\em nCOP}, a method that
incorporates individual mutational profiles with protein--protein
interaction networks, is a powerful approach for uncovering cancer
genes. Our method is based on an intuitive mathematical formulation
and demonstrates higher precision than other state-of-the-art methods
in detecting known cancer genes. Further, our approach is particularly beneficial in highlighting infrequently
mutated genes that are nevertheless relevant for cancer. Our approach therefore
complements existing frequency-based methods
(e.g.,~\cite{lawrence2013mutational}) that generally rely on comparisons to
background mutational models and lack the statistical power to detect
genes mutated in fewer individuals.

\begin{floatingfigure}[t]{0.4\linewidth}
\includegraphics[width=0.35\textwidth]{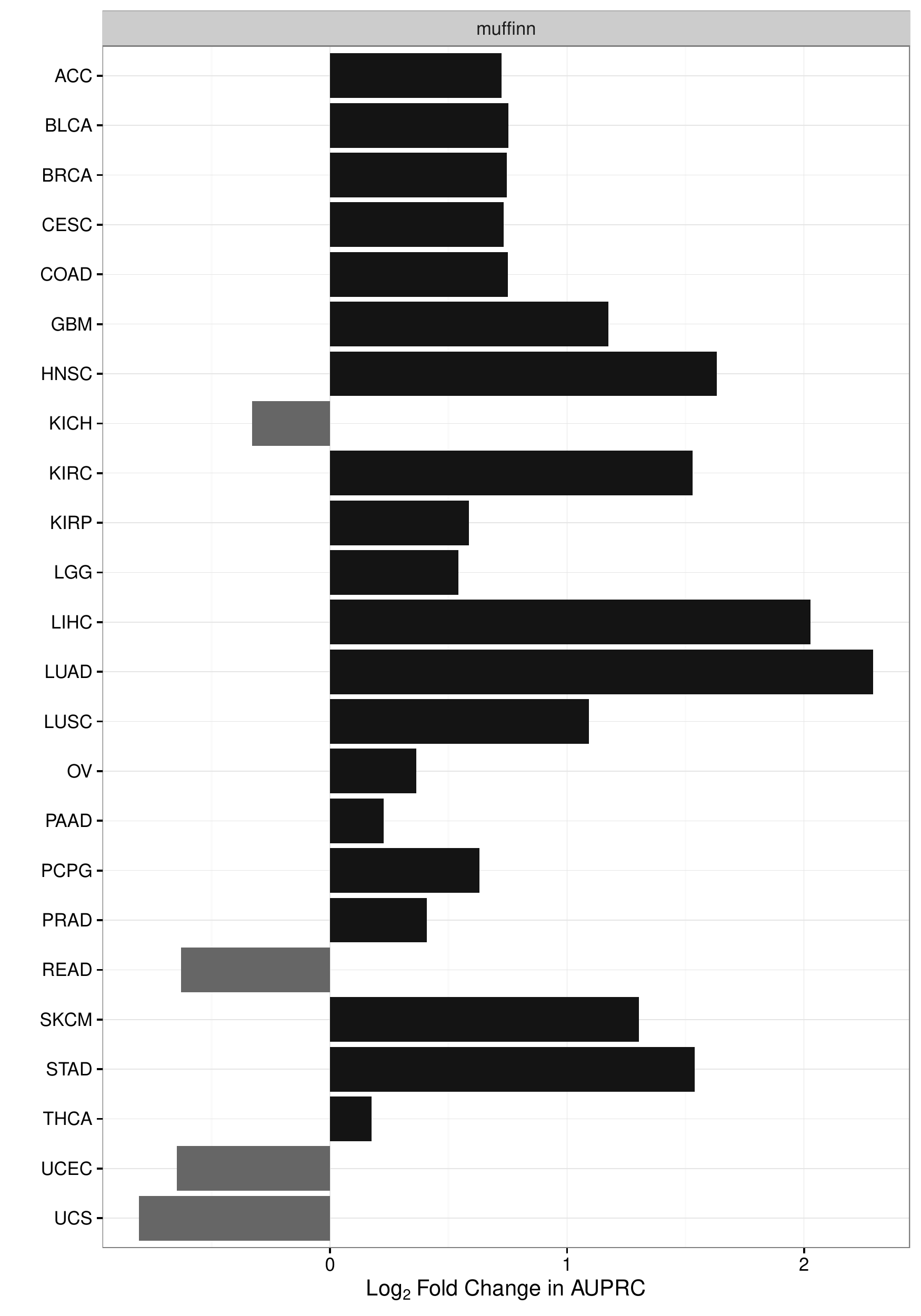}
\caption{\small {\bf Comparison of {\tt nCOP} to {\tt Muffinn}.}  For
  each of the 24 cancers, we compute the $log_2$ ratio of {\tt nCOP}'s
  to {\tt Muffinn}'s AUPRCs for their top 100 predictions. Our
  approach {\tt nCOP} outperforms  {\tt Muffinn}, a recent network-based approach, on
  20 out of 24 cancer types.}
\end{floatingfigure}

In the future, {\em nCOP} can be extended in a number of natural ways.
First, while {\em nCOP} currently analyzes only mutations within genes
that affect protein coding, other alterations are also commonly
observed in cancers. For example, copy number variants (CNVs) are
found frequently in cancers and can play critical functional roles
~\cite{ZackScCa13}. Although {\tt nCOP} does not currently use CNV
information, our framework can be extended to incorporate this
data. Indeed, as the numbers of CNVs and point mutations found within
each cancer genome appear to be inversely
related~\cite{CirielloMiAk13}, considering both types of alterations
will increase the power of our approach.  Second, {\em nCOP} may also
benefit from incorporating gene weights that reflect its likelihood to
play a role in cancer; in our current work, we consider a gene's
length but no other gene-specific attributes are considered.  Such
gene weights may be derived from existing approaches to detect
frequency of mutation or to assess the functional impact of mutations.
Finally, while {\tt nCOP} can output groups of genes that are not part
of a single connected component due to our randomized aggregation
procedure, extending {\tt nCOP}'s core algorithms to explicitly
consider multiple subnetworks corresponding to distinct pathways may
be a particularly promising avenue for future work.

We conclude by noting that researchers can use our framework to
rapidly and easily prioritize cancer genes, as {\em nCOP} requires
only straightforward inputs and runs on a desktop machine. Indeed,
{\tt nCOP}'s efficiency, robustness, and ease of use make it an excellent
choice to investigate cancer as well as possibly other
complex diseases. As sequencing costs plummet and cancer and other
disease sequencing mutational data become more abundant, the
predictive power of our method should only increase. In sum, we expect
that our method {\tt nCOP} will be of broad utility, and will represent a
valuable resource for the cancer community.

\newpage

\small
\bibliography{local}

\begin{thebibliography}{46}
\providecommand{\natexlab}[1]{#1}
\providecommand{\url}[1]{\texttt{#1}}
\expandafter\ifx\csname urlstyle\endcsname\relax
  \providecommand{\doi}[1]{doi: #1}\else
  \providecommand{\doi}{doi: \begingroup \urlstyle{rm}\Url}\fi

\bibitem[Babaei et~al.(2013)Babaei, Hulsman, Reinders, and
  de~Ridder]{babaei2013}
S.~Babaei, M.~Hulsman, M.~Reinders, and J.~de~Ridder.
\newblock Detecting recurrent gene mutation in interaction network context
  using multi-scale graph diffusion.
\newblock \emph{BMC Bioinformatics}, 14:\penalty0 29, 2013.

\bibitem[Bozic et~al.(2010)Bozic, Antal, Ohtsuki, Carter, Kim, Chen, Karchin,
  Kinzler, Vogelstein, and Nowak]{Bozic2010}
I.~Bozic, T.~Antal, H.~Ohtsuki, H.~Carter, D.~Kim, S.~Chen, R.~Karchin, K.~W.
  Kinzler, B.~Vogelstein, and M.~A. Nowak.
\newblock Accumulation of driver and passenger mutations during tumor
  progression.
\newblock \emph{Proc Natl Acad Sci U S A}, 107\penalty0 (43):\penalty0
  18545--50, 2010.

\bibitem[Cao et~al.(2013)Cao, Zhang, Park, Daniels, Crovella, Cowen, and
  Hescott]{cao2013going}
M.~Cao, H.~Zhang, J.~Park, N.~M. Daniels, M.~E. Crovella, L.~J. Cowen, and
  B.~Hescott.
\newblock Going the distance for protein function prediction: a new distance
  metric for protein interaction networks.
\newblock \emph{PloS one}, 8\penalty0 (10):\penalty0 e76339, 2013.

\bibitem[Cerami et~al.(2010)Cerami, Demir, Schultz, Taylor, and
  Sander]{Netbox2010}
E.~Cerami, E.~Demir, N.~Schultz, B.~S. Taylor, and C.~Sander.
\newblock {{A}utomated network analysis identifies core pathways in
  glioblastoma}.
\newblock \emph{PLoS ONE}, 5\penalty0 (2):\penalty0 e8918, 2010.

\bibitem[Cho et~al.(2016)Cho, Shim, Kim, Supek, Lehner, and
  Lee]{cho2016muffinn}
A.~Cho, J.~E. Shim, E.~Kim, F.~Supek, B.~Lehner, and I.~Lee.
\newblock Muffinn: cancer gene discovery via network analysis of somatic
  mutation data.
\newblock \emph{Genome Biology}, 17\penalty0 (1):\penalty0 129, 2016.

\bibitem[Chowdhury and Koyuturk(2010)]{ChowdhuryKo10}
S.~Chowdhury and M.~Koyuturk.
\newblock Identification of coordinately dysregulated subnetworks in complex
  phenotypes.
\newblock In \emph{Pac Symp Biocomput}, pages 133–--144, 2010.

\bibitem[Ciriello et~al.(2012)Ciriello, Cerami, Sander, and Schultz]{memo2012}
G.~Ciriello, E.~Cerami, C.~Sander, and N.~Schultz.
\newblock {{M}utual exclusivity analysis identifies oncogenic network modules}.
\newblock \emph{Genome Res.}, 22\penalty0 (2):\penalty0 398--406, Feb 2012.

\bibitem[Ciriello et~al.(2013)Ciriello, Miller, Aksoy, Senbabaoglu, Schultz,
  and Sander]{CirielloMiAk13}
G.~Ciriello, M.~Miller, B.~Aksoy, Y.~Senbabaoglu, N.~Schultz, and C.~Sander.
\newblock Emerging landscape of oncogenic signatures across human cancers.
\newblock \emph{Nature Genetics}, 45:\penalty0 1127--1133, 2013.

\bibitem[Dand et~al.(2013)Dand, Sprengel, Ahlers, and
  Schlitt]{dand2013biogranat}
N.~Dand, F.~Sprengel, V.~Ahlers, and T.~Schlitt.
\newblock {BioGranat-IG}: a network analysis tool to suggest mechanisms of
  genetic heterogeneity from exome-sequencing data.
\newblock \emph{Bioinformatics}, 29\penalty0 (6):\penalty0 733--741, 2013.

\bibitem[Dees et~al.(2012)Dees, Zhang, Kandoth, Wendl, Schierding, Koboldt,
  et~al.]{DeesZhKa12}
N.~Dees, Q.~Zhang, C.~Kandoth, M.~Wendl, W.~Schierding, D.~Koboldt, et~al.
\newblock Mu{S}i{C}: Identifying mutational significance in cancer genomes.
\newblock \emph{Genome Res.}, 22:\penalty0 1589–--1598, 2012.

\bibitem[Even and Tarjan(1975)]{even1975network}
S.~Even and R.~E. Tarjan.
\newblock Network flow and testing graph connectivity.
\newblock \emph{SIAM journal on computing}, 4\penalty0 (4):\penalty0 507--518,
  1975.

\bibitem[Forbes et~al.(2010)Forbes, Bindal, Bamford, Cole, Kok, Beare, Jia,
  Shepherd, Leung, Menzies, et~al.]{forbes2010cosmic}
S.~A. Forbes, N.~Bindal, S.~Bamford, C.~Cole, C.~Y. Kok, D.~Beare, M.~Jia,
  R.~Shepherd, K.~Leung, A.~Menzies, et~al.
\newblock Cosmic: mining complete cancer genomes in the catalogue of somatic
  mutations in cancer.
\newblock \emph{Nucleic acids research}, page gkq929, 2010.

\bibitem[Futreal et~al.(2004)Futreal, Coin, Marshall, Down, Hubbard, Wooster,
  Rahman, and Stratton]{Futreal2004}
P.~A. Futreal, L.~Coin, M.~Marshall, T.~Down, T.~Hubbard, R.~Wooster,
  N.~Rahman, and M.~R. Stratton.
\newblock A census of human cancer genes.
\newblock \emph{Nat Rev Cancer}, 4\penalty0 (3):\penalty0 177--83, 2004.

\bibitem[Garraway and Lander(2013)]{Garraway2013}
L.~A. Garraway and E.~S. Lander.
\newblock Lessons from the cancer genome.
\newblock \emph{Cell}, 153\penalty0 (1):\penalty0 17--37, 2013.

\bibitem[Hartwell et~al.(1999)Hartwell, Hopfield, Leibler, and
  Murray]{HartwellHoLe99}
L.~Hartwell, J.~Hopfield, S.~Leibler, and A.~Murray.
\newblock From molecular to modular cell biology.
\newblock \emph{Nature}, 402:\penalty0 C47--52, 1999.

\bibitem[Hastie et~al.(2009)Hastie, Tibshirani, Friedman, Hastie, Friedman, and
  Tibshirani]{hastie2009elements}
T.~Hastie, R.~Tibshirani, J.~Friedman, T.~Hastie, J.~Friedman, and
  R.~Tibshirani.
\newblock \emph{The elements of statistical learning}, volume~2.
\newblock Springer, 2009.

\bibitem[Hofree et~al.(2013)Hofree, Shen, Carter, Gross, and
  Ideker]{hofree2013network}
M.~Hofree, J.~P. Shen, H.~Carter, A.~Gross, and T.~Ideker.
\newblock Network-based stratification of tumor mutations.
\newblock \emph{Nature methods}, 2013.

\bibitem[Hudson et~al.(2010)Hudson, Anderson, Aretz, Barker, Bell, Bernab{\'e},
  Bhan, Calvo, Eerola, Gerhard, et~al.]{hudson2010international}
T.~J. Hudson, W.~Anderson, A.~Aretz, A.~D. Barker, C.~Bell, R.~R. Bernab{\'e},
  M.~Bhan, F.~Calvo, I.~Eerola, D.~S. Gerhard, et~al.
\newblock International network of cancer genome projects.
\newblock \emph{Nature}, 464\penalty0 (7291):\penalty0 993--998, 2010.

\bibitem[ILOG CPLEX()]{CPLEX}
ILOG CPLEX.
\newblock {ILOG CPLEX} 7.1, 2016.
\newblock http://www.ilog.com/products/cplex/.

\bibitem[Jia and Zhao(2014)]{jia2014varwalker}
P.~Jia and Z.~Zhao.
\newblock Varwalker: personalized mutation network analysis of putative cancer
  genes from next-generation sequencing data.
\newblock \emph{PLoS Comput Biol}, 10\penalty0 (2):\penalty0 e1003460, 2014.

\bibitem[Jones et~al.(2008)Jones, Zhang, Parsons, Lin, Leary, Angenendt,
  Mankoo, Carter, Kamiyama, Jimeno, et~al.]{jones2008core}
S.~Jones, X.~Zhang, D.~W. Parsons, J.~C.-H. Lin, R.~J. Leary, P.~Angenendt,
  P.~Mankoo, H.~Carter, H.~Kamiyama, A.~Jimeno, et~al.
\newblock Core signaling pathways in human pancreatic cancers revealed by
  global genomic analyses.
\newblock \emph{Science Signaling}, 321\penalty0 (5897):\penalty0 1801, 2008.

\bibitem[Katz et~al.(1975)Katz, Emmanouel, and Lindheimer]{katz1975thyroid}
A.~Katz, D.~Emmanouel, and M.~Lindheimer.
\newblock Thyroid hormone and the kidney.
\newblock \emph{Nephron}, 15\penalty0 (3-5):\penalty0 223--249, 1975.

\bibitem[Kim et~al.(2011)Kim, Wuchty, and Przytycka]{KimWuPr11}
Y.~Kim, S.~Wuchty, and T.~Przytycka.
\newblock Identifying causal genes and dysregulated pathways in complex
  diseases.
\newblock \emph{PLoS Comput Biol}, 7:\penalty0 e1001095, 2011.

\bibitem[Kim et~al.(2015)Kim, Cho, Dao, and Przytycka]{kim2015memcover}
Y.-A. Kim, D.-Y. Cho, P.~Dao, and T.~M. Przytycka.
\newblock Memcover: integrated analysis of mutual exclusivity and functional
  network reveals dysregulated pathways across multiple cancer types.
\newblock \emph{Bioinformatics}, 31\penalty0 (12):\penalty0 i284--i292, 2015.

\bibitem[Lawrence et~al.(2013)Lawrence, Stojanov, Polak, Kryukov, Cibulskis,
  Sivachenko, Carter, Stewart, Mermel, Roberts, et~al.]{lawrence2013mutational}
M.~S. Lawrence, P.~Stojanov, P.~Polak, G.~V. Kryukov, K.~Cibulskis,
  A.~Sivachenko, S.~L. Carter, C.~Stewart, C.~H. Mermel, S.~A. Roberts, et~al.
\newblock Mutational heterogeneity in cancer and the search for new
  cancer-associated genes.
\newblock \emph{Nature}, 499\penalty0 (7457):\penalty0 214--218, 2013.

\bibitem[Leiserson et~al.(2015)Leiserson, Vandin, Wu, Dobson, Eldridge, Thomas,
  Papoutsaki, Kim, Niu, McLellan, Lawrence, Gonzalez-Perez, Tamborero, Cheng,
  Ryslik, Lopez-Bigas, Getz, Ding, and Raphael]{Leiserson15}
M.~D.~M. Leiserson, F.~Vandin, H.-T. Wu, J.~R. Dobson, J.~V. Eldridge, J.~L.
  Thomas, A.~Papoutsaki, Y.~Kim, B.~Niu, M.~McLellan, M.~S. Lawrence,
  A.~Gonzalez-Perez, D.~Tamborero, Y.~Cheng, G.~A. Ryslik, N.~Lopez-Bigas,
  G.~Getz, L.~Ding, and B.~J. Raphael.
\newblock Pan-cancer network analysis identifies combinations of rare somatic
  mutations across pathways and protein complexes.
\newblock \emph{Nature Genetics}, 47:\penalty0 106--114, 2015.

\bibitem[Lin et~al.(2014)Lin, Aihara, Chung, Li, Huang, Chen, Weng, Carlson,
  Wands, and Dong]{lin2014lrh1}
Q.~Lin, A.~Aihara, W.~Chung, Y.~Li, Z.~Huang, X.~Chen, S.~Weng, R.~I. Carlson,
  J.~R. Wands, and X.~Dong.
\newblock Lrh1 as a driving factor in pancreatic cancer growth.
\newblock \emph{Cancer letters}, 345\penalty0 (1):\penalty0 85--90, 2014.

\bibitem[McLendon et~al.(2008)McLendon, Friedman, Bigner, Van~Meir, Brat,
  Mastrogianakis, Olson, Mikkelsen, Lehman, Aldape,
  et~al.]{mclendon2008comprehensive}
R.~McLendon, A.~Friedman, D.~Bigner, E.~G. Van~Meir, D.~J. Brat, G.~M.
  Mastrogianakis, J.~J. Olson, T.~Mikkelsen, N.~Lehman, K.~Aldape, et~al.
\newblock Comprehensive genomic characterization defines human glioblastoma
  genes and core pathways.
\newblock \emph{Nature}, 455\penalty0 (7216):\penalty0 1061--1068, 2008.

\bibitem[Paull et~al.(2013)Paull, Carlin, Niepel, Sorger, Haussler, and
  Stuart]{tiedie2013}
E.~O. Paull, D.~E. Carlin, M.~Niepel, P.~K. Sorger, D.~Haussler, and J.~M.
  Stuart.
\newblock {{D}iscovering causal pathways linking genomic events to
  transcriptional states using {T}ied {D}iffusion {T}hrough {I}nteracting
  {E}vents ({T}ie{D}{I}{E})}.
\newblock \emph{Bioinformatics}, 29\penalty0 (21):\penalty0 2757--2764, Nov
  2013.

\bibitem[Prasad et~al.(2009)Prasad, Goel, Kandasamy, Keerthikumar, Kumar,
  Mathivanan, Telikicherla, Raju, Shafreen, Venugopal, et~al.]{prasad2009human}
T.~K. Prasad, R.~Goel, K.~Kandasamy, S.~Keerthikumar, S.~Kumar, S.~Mathivanan,
  D.~Telikicherla, R.~Raju, B.~Shafreen, A.~Venugopal, et~al.
\newblock Human protein reference database—2009 update.
\newblock \emph{Nucleic acids research}, 37\penalty0 (suppl 1):\penalty0
  D767--D772, 2009.

\bibitem[Shuai and Hu(2006)]{shuai2006connected}
T.-P. Shuai and X.-D. Hu.
\newblock Connected set cover problem and its applications.
\newblock In \emph{International Conference on Algorithmic Applications in
  Management}, pages 243--254. Springer, 2006.

\bibitem[Spirin and Mirny(2003)]{SpirinMi03}
V.~Spirin and L.~A. Mirny.
\newblock Protein complexes and functional modules in molecular networks.
\newblock \emph{Proc. Natl. Acad. Sci. USA.}, 100:\penalty0 12123--12128, 2003.

\bibitem[Stark et~al.(2006)Stark, Breitkreutz, Reguly, Boucher, Breitkreutz,
  and Tyers]{stark2006biogrid}
C.~Stark, B.-J. Breitkreutz, T.~Reguly, L.~Boucher, A.~Breitkreutz, and
  M.~Tyers.
\newblock Biogrid: a general repository for interaction datasets.
\newblock \emph{Nucleic acids research}, 34\penalty0 (suppl 1):\penalty0
  D535--D539, 2006.

\bibitem[Stratton et~al.(2009)Stratton, Campbell, and Futreal]{Stratton2009}
M.~R. Stratton, P.~J. Campbell, and P.~A. Futreal.
\newblock The cancer genome.
\newblock \emph{Nature}, 458\penalty0 (7239):\penalty0 719--24, 2009.

\bibitem[{TCGA Research Network:}()]{tcga}
{TCGA Research Network:}.
\newblock {http://cancergenome.nih.gov/}.

\bibitem[{The Cancer Genome Atlas Network}(2012)]{TCGA-colorectal}
{The Cancer Genome Atlas Network}.
\newblock Comprehensive molecular characterization of human colon and rectal
  cancer.
\newblock \emph{Nature}, 487:\penalty0 330--337, 2012.

\bibitem[{The {I}nternational {C}ancer {G}enome {C}onsortium}(2010)]{ICGC}
{The {I}nternational {C}ancer {G}enome {C}onsortium}.
\newblock International network of cancer genome projects.
\newblock \emph{Nature}, 464:\penalty0 993--998, 2010.

\bibitem[Ulitsky et~al.(2010)Ulitsky, Krishnamurthy, Karp, and
  Shamir]{ulitsky2010degas}
I.~Ulitsky, A.~Krishnamurthy, R.~M. Karp, and R.~Shamir.
\newblock Degas: de novo discovery of dysregulated pathways in human diseases.
\newblock \emph{PLoS one}, 5\penalty0 (10):\penalty0 e13367, 2010.

\bibitem[Vandin et~al.(2011)Vandin, Upfal, and Raphael]{vandin2011algorithms}
F.~Vandin, E.~Upfal, and B.~J. Raphael.
\newblock Algorithms for detecting significantly mutated pathways in cancer.
\newblock \emph{Journal of Computational Biology}, 18\penalty0 (3):\penalty0
  507--522, 2011.

\bibitem[Vaske et~al.(2010)Vaske, Benz, Sanborn, Earl, Szeto, Zhu, Haussler,
  and Stuart]{VaskeBeSa10}
C.~J. Vaske, S.~C. Benz, J.~Z. Sanborn, D.~Earl, C.~Szeto, J.~Zhu, D.~Haussler,
  and J.~M. Stuart.
\newblock {{I}nference of patient-specific pathway activities from
  multi-dimensional cancer genomics data using {P}{A}{R}{A}{D}{I}{G}{M}}.
\newblock \emph{Bioinformatics}, 26\penalty0 (12):\penalty0 i237--245, Jun
  2010.

\bibitem[Vogelstein et~al.(2013)Vogelstein, Papadopoulos, Velculescu, Zhou,
  Diaz, and Kinzler]{Vogelstein2013}
B.~Vogelstein, N.~Papadopoulos, V.~E. Velculescu, S.~Zhou, J.~Diaz, L.~A., and
  K.~W. Kinzler.
\newblock Cancer genome landscapes.
\newblock \emph{Science}, 339\penalty0 (6127):\penalty0 1546--58, 2013.

\bibitem[Wendl et~al.(2011)Wendl, Wallis, Lin, Kandoth, Mardis, Wilson, and
  Ding]{pmid21498403}
M.~C. Wendl, J.~W. Wallis, L.~Lin, C.~Kandoth, E.~R. Mardis, R.~K. Wilson, and
  L.~Ding.
\newblock {{P}ath{S}can: a tool for discerning mutational significance in
  groups of putative cancer genes}.
\newblock \emph{Bioinformatics}, 27\penalty0 (12):\penalty0 1595--1602, Jun
  2011.

\bibitem[Wolf et~al.(2014)Wolf, M{\"u}ller-Decker, Flechtenmacher, Zhang,
  Shahmoradgoli, Mills, Hoheisel, and Boettcher]{wolf2014vivo}
J.~Wolf, K.~M{\"u}ller-Decker, C.~Flechtenmacher, F.~Zhang, M.~Shahmoradgoli,
  G.~Mills, J.~Hoheisel, and M.~Boettcher.
\newblock An in vivo rnai screen identifies sall1 as a tumor suppressor in
  human breast cancer with a role in cdh1 regulation.
\newblock \emph{Oncogene}, 33\penalty0 (33):\penalty0 4273--4278, 2014.

\bibitem[Youn and Simon(2011)]{Youn2011}
A.~Youn and R.~Simon.
\newblock Identifying cancer driver genes in tumor genome sequencing studies.
\newblock \emph{Bioinformatics}, 27\penalty0 (2), 2011.

\bibitem[Yu et~al.(2016)Yu, Li, and Shen]{yu2016brd7}
X.~Yu, Z.~Li, and J.~Shen.
\newblock Brd7: a novel tumor suppressor gene in different cancers.
\newblock \emph{American journal of translational research}, 8\penalty0
  (2):\penalty0 742, 2016.

\bibitem[Zack et~al.(2013)Zack, Schumacher, Carter, Cherniack, Saksena, Tabak,
  et~al.]{ZackScCa13}
T.~I. Zack, S.~E. Schumacher, S.~L. Carter, A.~D. Cherniack, G.~Saksena,
  B.~Tabak, et~al.
\newblock {{P}an-cancer patterns of somatic copy number alteration}.
\newblock \emph{Nat. Genet.}, 45\penalty0 (10):\penalty0 1134--1140, Oct 2013.

\end{thebibliography}
\bibliographystyle{abbrvnat}
\newpage



\noindent
{\large
\section*{Supplementary Tables and Figures}
}
\bigskip

\noindent
The following pages contain a table summarizing the TCGA data we use
along with 6 supplementary figures that support the findings of the
main paper.

\newpage
\beginsupplement

\begin{table}[t!]
\includegraphics[width=1\textwidth]{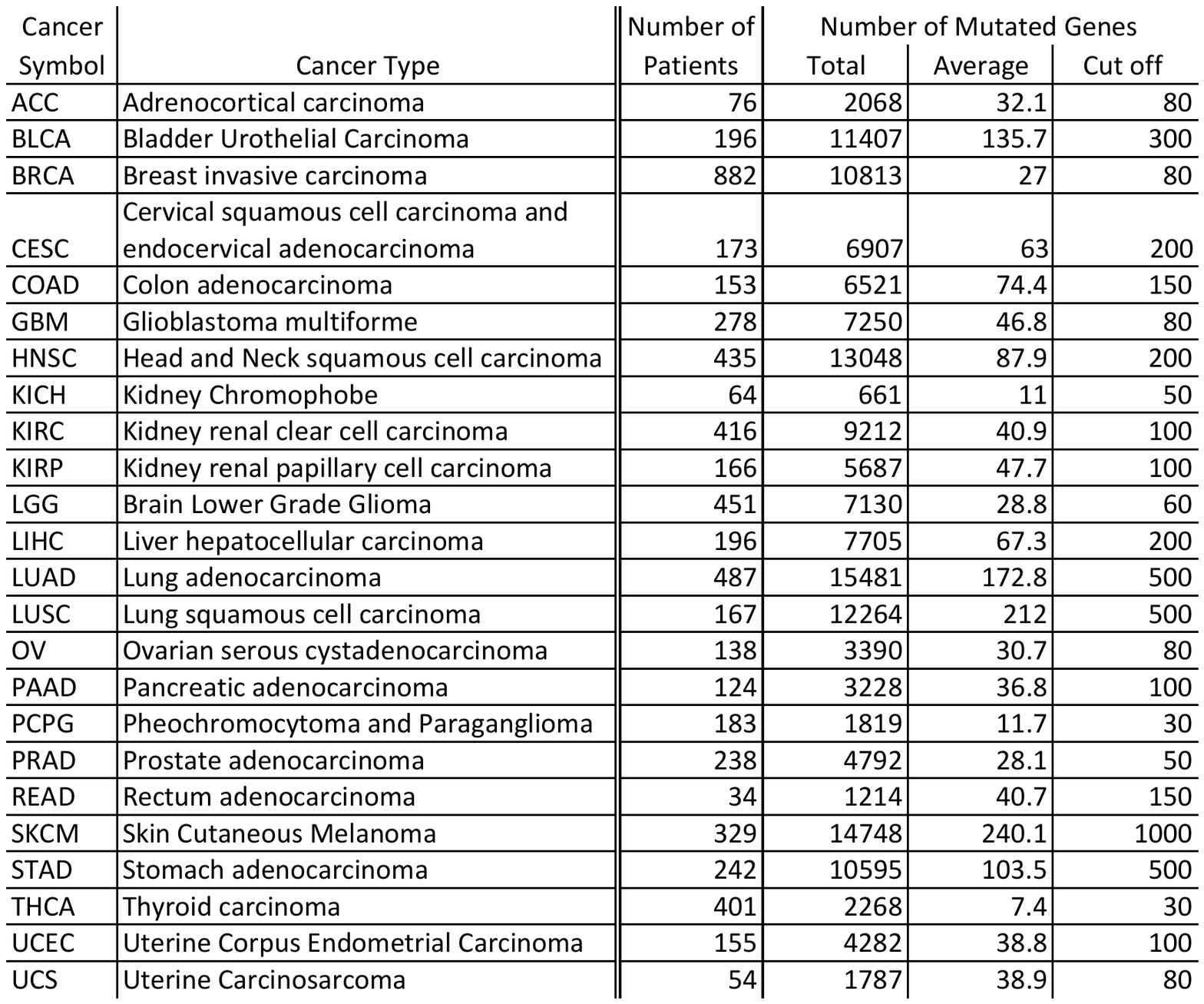}
\vspace{-3in}
\caption{{\bf TCGA dataset and statistics.} We list the 24 cancer types studied
  along with their abbreviations. For each cancer type, we give the
  total number of patient samples considered after highly mutated
  samples are filtered out, the total number of mutated genes across
  these samples, the average number of mutated genes across
  all samples, and the cutoff on the number of mutated genes
  within a sample that was used to filter samples.}
\end{table}

\clearpage
\newpage

\begin{figure}[H]
\includegraphics[width=1\textwidth]{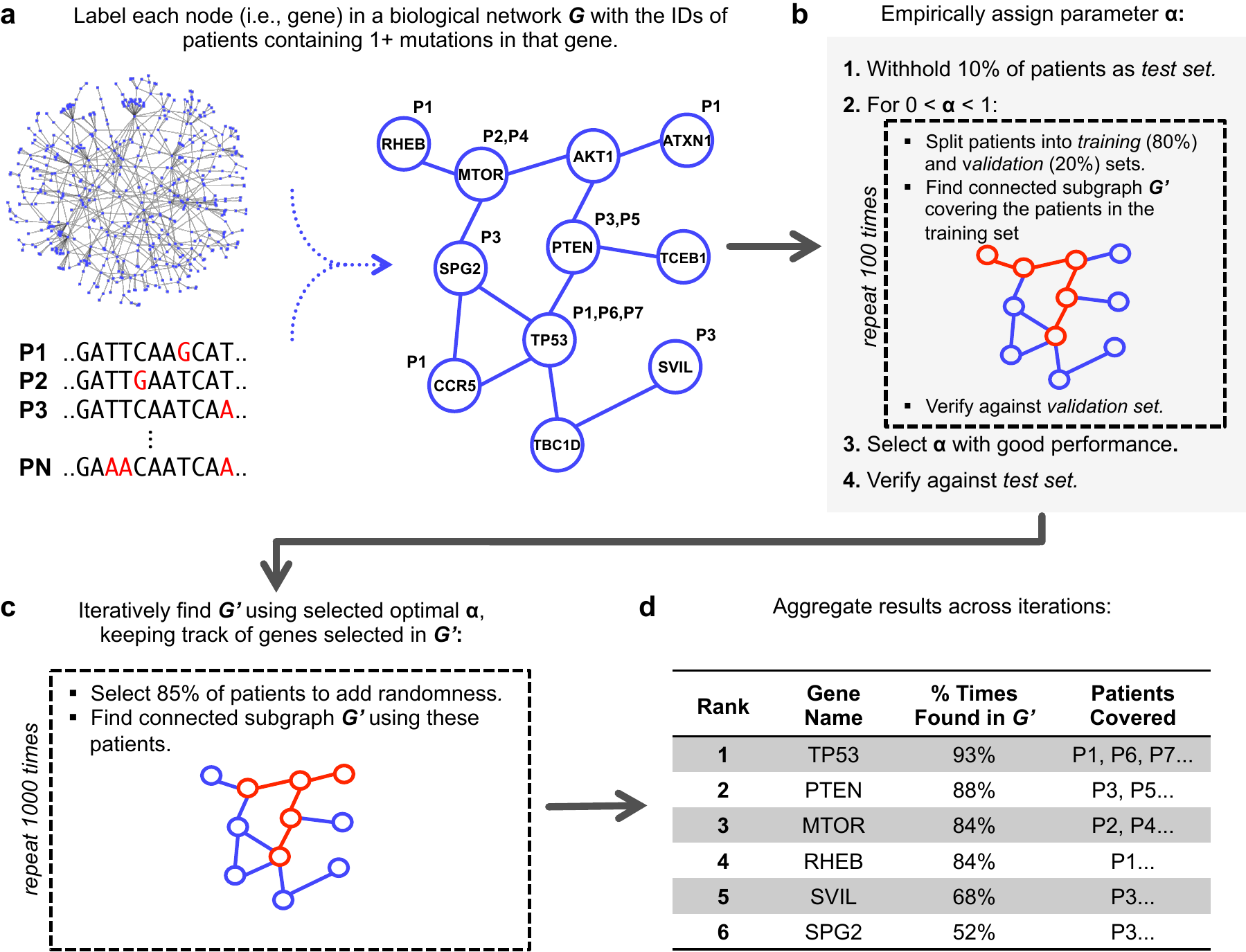}

\caption{\small {\bf Overview of our approach.} {\bf (a)} Somatic mutations
  are mapped onto a protein-protein interaction network.  Each node is
  associated with the set of individuals whose cancers have mutations
  in the corresponding gene. The overall goal is to select a small
  connected subnetwork such that most individuals in the cohort have
  mutations in one of the corresponding genes (i.e., are
  ``covered''). {\bf (b)} {\tt nCOP} automatically selects a value for the
  parameter $\alpha$ by performing a series of cross-validation
  tests. First, $10\%$ of the individuals are withheld as a test
  set. Next, the remaining individuals are
  repeatedly and randomly split into two groups, a training set
  ($80\%$) and a validation set ($20\%$). For each split, the {\tt nCOP}
  search heuristic is run for a range of $\alpha$ values ($0 < \alpha < 1$)  using the individuals comprising the
  training set. The parameter $\alpha$ is selected to obtain high
  coverage of the individuals in the validation sets while maintaining
  similar coverage on the training sets (i.e., not overfitting to the
  training set). Coverage of individuals in the initially withheld
  test set is also calculated and confirmed to be similar to the validation sets. {\bf (c)} Once $\alpha$ is selected, to
  avoid overfitting on the entire dataset, {\tt nCOP} is run 1000 times
  using $85\%$ of the individuals. {\bf (d)} Finally, the subnetworks
  output across the runs are aggregated and candidate genes are ranked
  by the number of the times they appear in the outputted
  subnetworks. }

\end{figure}
\clearpage
\newpage


\begin{figure}[t!]
\includegraphics[width=\textwidth,height=\textheight,keepaspectratio]{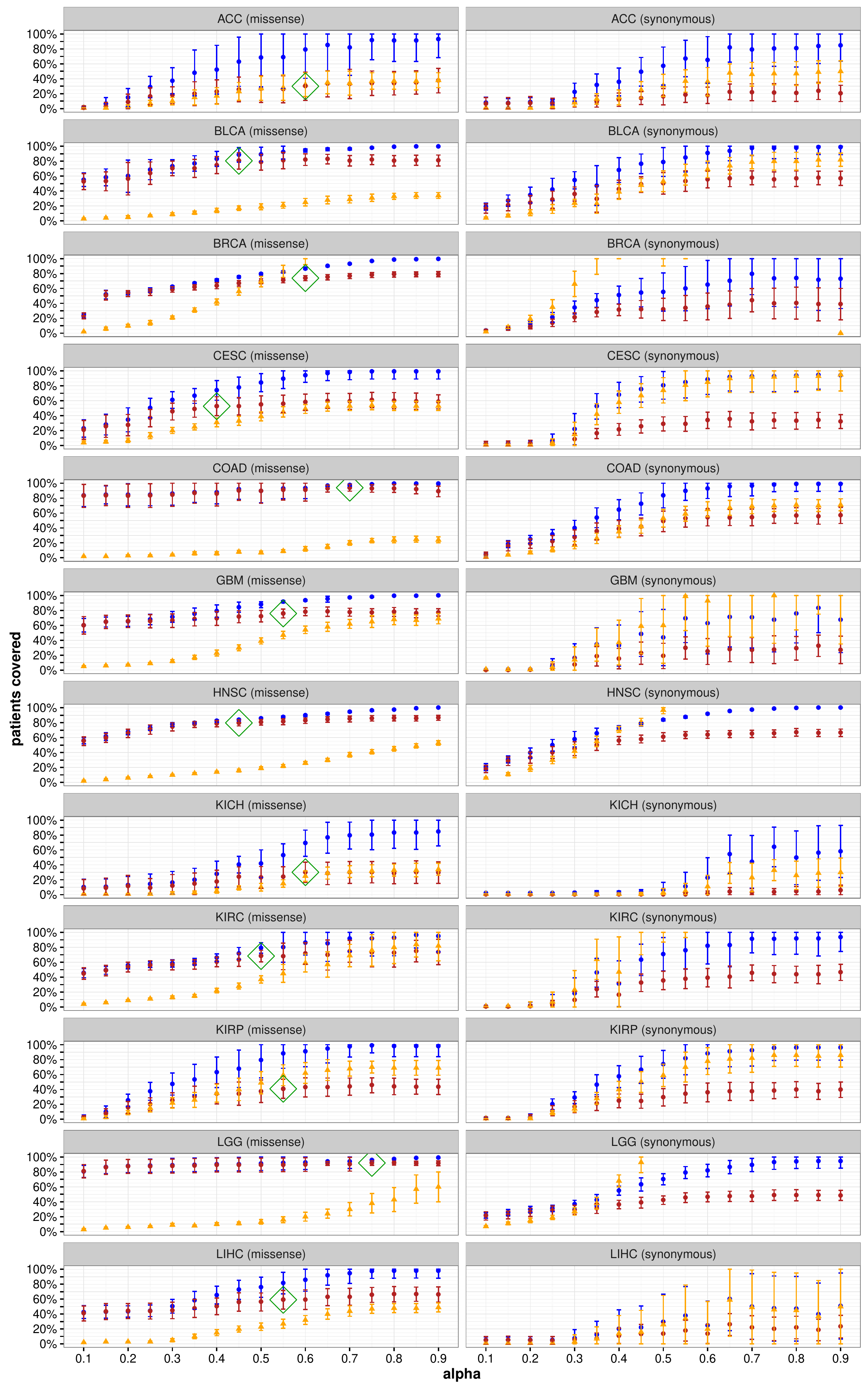}
\end{figure}

\begin{figure}[t!]
\includegraphics[width=\textwidth,height=\textheight,keepaspectratio]{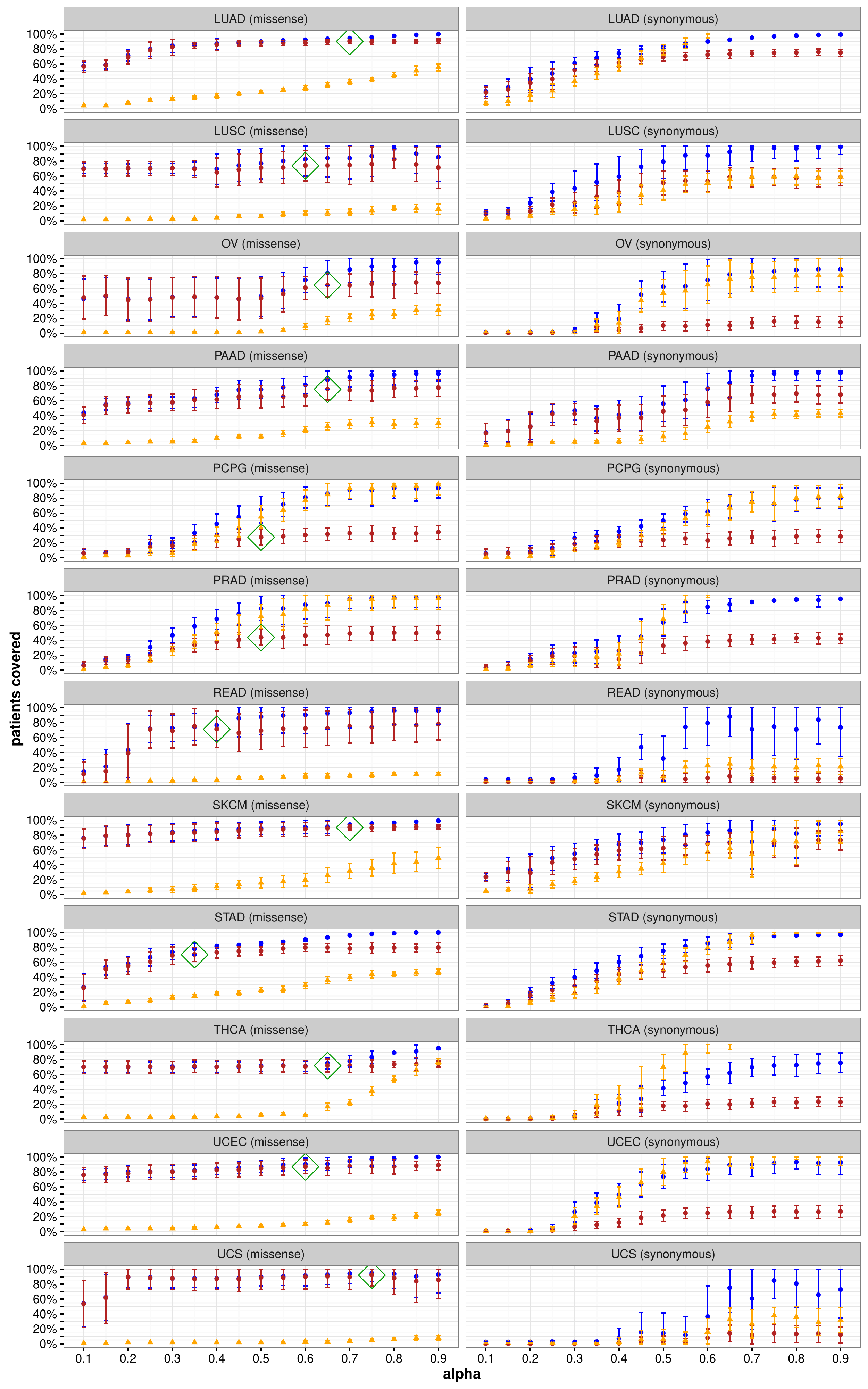}
\end{figure}


\clearpage
\newpage

\captionof{figure}{{\bf Fraction of individuals covered as $\alpha$
    varies across all cancers.}  For each random split of the
  individuals, we run our algorithm on the training sets for different
  values of $\alpha$, and plot the fraction of covered individuals in
  the training (blue) and validation (red) sets.  We also give the
  number of proteins in the uncovered subgraphs $G'$ (orange). For each plotted
  value, the mean and standard deviation over 100 random splits are
  shown and the automatically selected $\alpha$ for the missense mutation
  data is indicated by a green rhombus. The performances on both the
  training and validation sets are much worse when using synonymous
  mutations compared to when using missense mutations. Coverage
  on the validation set for synonymous mutations is consistently lower
  for the same values of $\alpha$ across respective cancer types than
  that for missense mutations, with maximum possible coverage on the
  validation set not exceeding 50\% in many cases. Further, it takes
  significantly more nodes to cover the same fraction of patients when
  using synonymous mutations.}


\begin{figure}[t!]
\includegraphics[width=1\textwidth]{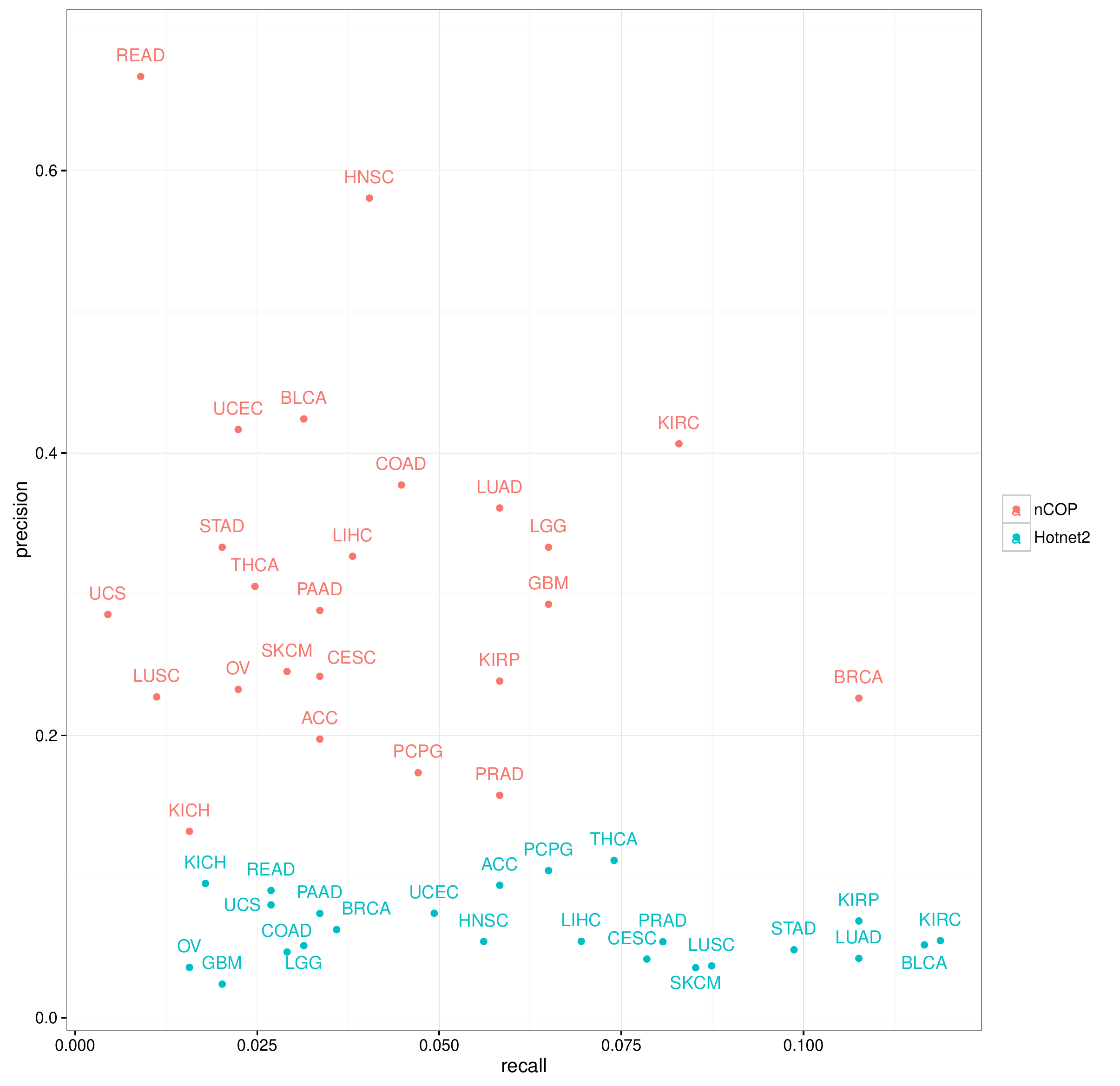}

\caption{{\bf Comparison between {\tt nCOP} and {\tt Hotnet2}.}  For
  each cancer type, we compute the precision and recall of the genes
  returned by {\tt nCOP} and {\tt Hotnet2}.   For {\tt nCOP}, we
  choose a single threshold to select predicted cancer genes,
  corresponding to those genes that occur in at least 25\% of the
  runs.  While {\tt Hotnet2} achieves slightly greater recall due to
  the larger number of genes it highlights, {\tt nCOP}'s precision
  using this threshold is superior. {\tt nCOP} uncovers fewer but
  potentially more relevant cancer genes. }
\end{figure}
\clearpage
\newpage

\begin{figure}[t!]
  \includegraphics[width=0.6\textwidth,keepaspectratio]{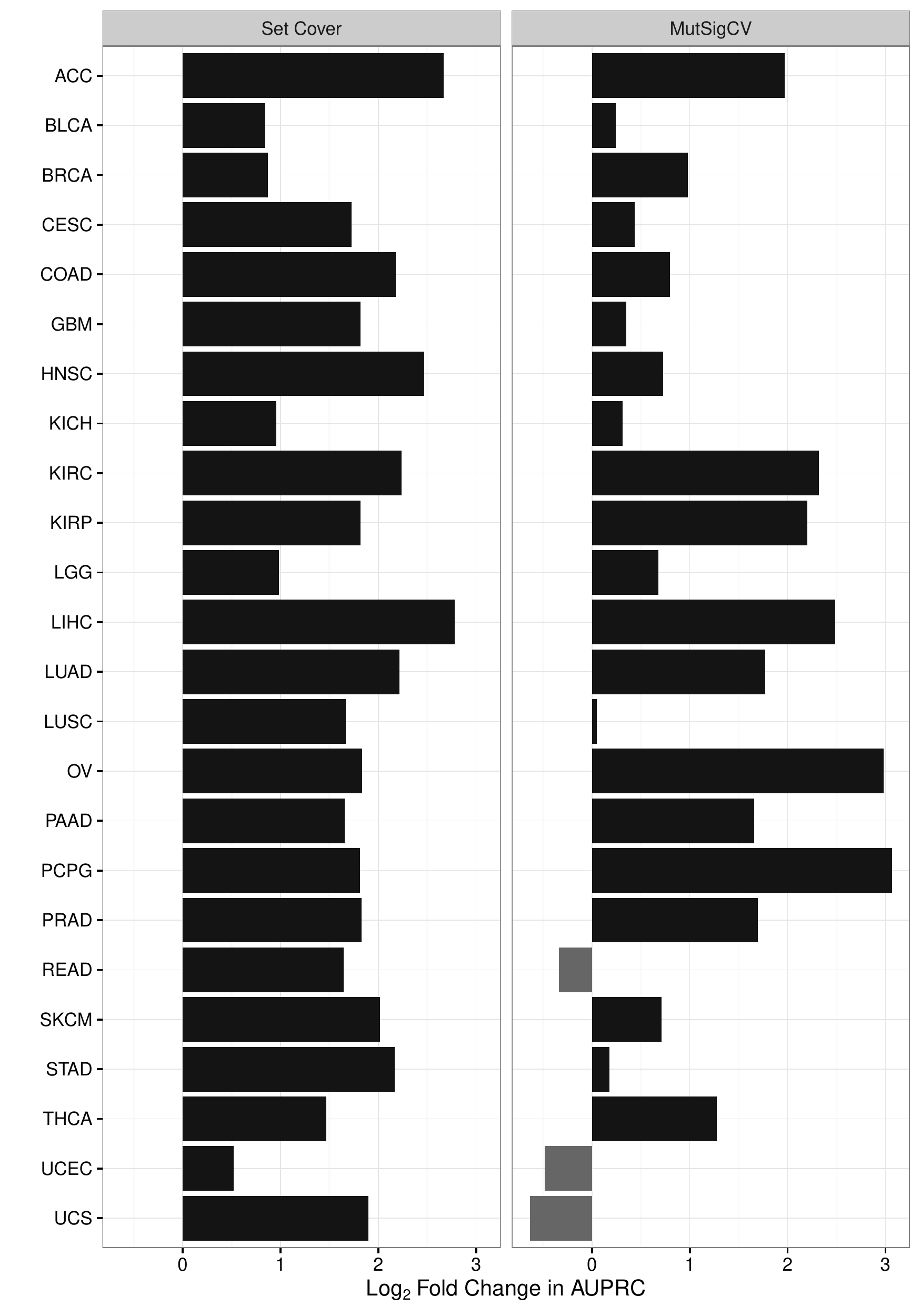}
  
  \caption{{\bf Comparison of \emph{{\tt nCOP}} when run on Biogrid to
      network-agnostic methods across 24 cancers.} To make sure that
    our method is robust with respect to the specific network utilized, we
    repeat our entire analysis procedure using the Biogrid network. For each of
    the 24 cancers, we compute the $\log_2$ ratio of AUPRCs using the top 100 predictions for
    \emph{{\tt nCOP}} and Set Cover (left panel) and \emph{{\tt nCOP}} and MutSigCV
    (right panel). Our approach \emph{{\tt nCOP}} outperforms the
    network-agnostic methods in 21  out of 24 of the cancer types.}

\end{figure}
\clearpage
\newpage


\begin{figure}[t!]
\includegraphics[width=0.6\textwidth,keepaspectratio]{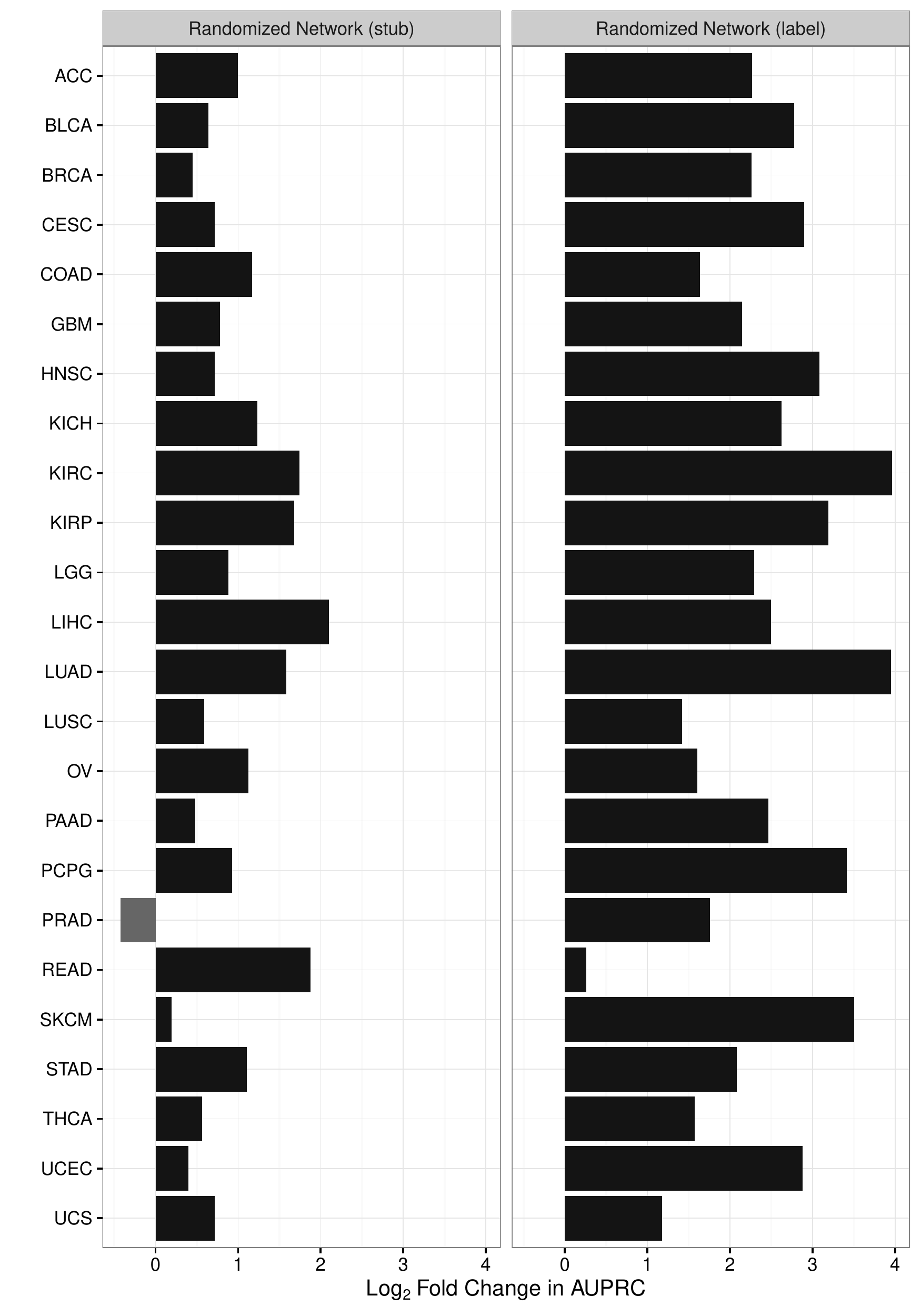}

\caption{{\bf Comparison to randomized networks.}  We use two
  approaches to randomize the underlying HPRD network: (1) a classic
  degree-preserving randomization (stub rewiring) and (2) node label
  shuffling where the network structure is maintained but genes can
  have different degrees and interactors.  For each of the 24 cancers,
  we compute the $log_2$ ratio of the area under the precision recall
  curve using {\tt nCOP} on the real network and on the randomized
  network.  Performance, as expected, is worse for both randomizations
  across all cancers, except, interestingly, for PRAD with stub
  rewiring. We speculate that the large number of CCG genes with high
  degree mutated in PRAD became connected in a module that was ranked
  highly by \emph{{\tt nCOP}}. In the full version of this paper, this
  figure will contain data averaged over multiple randomized networks.}

\end{figure}
\clearpage
\newpage


\begin{figure}[t!]
\includegraphics[width=1\textwidth]{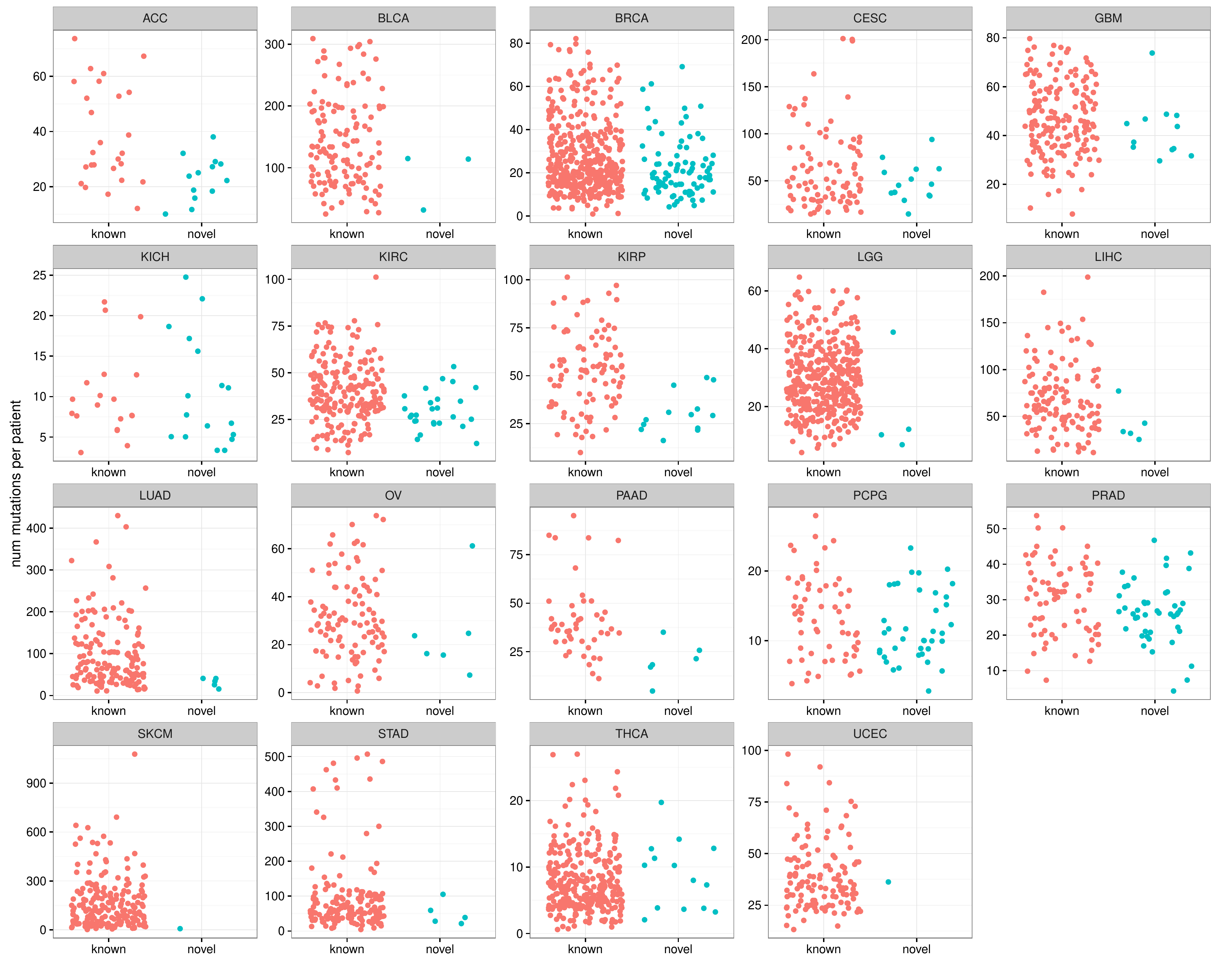}

\caption{{\bf Novel genes uncovered by \emph{{\tt nCOP}} are not due
    to patients with many mutations.}  Plotted for each cancer type
  are the total number of missense  mutations for patients
  having missense mutations only in known CCG genes and not in novel genes
  (left) and the total number of missense mutations for
  patients having missense mutations only in novel genes and not in CCG genes
  (right). The novel genes uncovered by \emph{{\tt nCOP}} are not due
  to patients with large numbers of mutations. }

\end{figure}
\clearpage
\newpage

\end{document}